\documentclass[prodmode]{acmsmall} 
\usepackage{amsmath}
\usepackage{graphicx}
\usepackage{listings}
\usepackage{natbib}
\usepackage{url}
\usepackage[justification=centering,font={footnotesize}]{caption}
\usepackage{ftnxtra}
\usepackage{threeparttable}
\usepackage{balance}
\usepackage{courier}
\usepackage[pdf]{pstricks}
\usepackage{pst-rel-points}
\usepackage{pst-node}
\usepackage{listings}
\usepackage{comment}
\usepackage{wrapfig}
\usepackage{tabularx}
\usepackage{floatrow}
\usepackage[inline]{enumitem}
\setlist{nolistsep}	
\usepackage{multirow}
\usepackage{tablefootnote}

\usepackage[titletoc,toc,title]{appendix}

\newcommand{\cmt}[1]{}

\newcommand{\TB}{\mbox{TB}}
\newcommand{\Release}{\textit{relssp}}

\newtheorem{condition}{Condition}
\newtheorem{definition}{Definition}[section]

\newcommand{\isfalse}{\mbox{false}}

\newcommand{\twofigure}[4]{%
  \begin{floatrow}
    \ffigbox{#1}{#2}
    \ffigbox{#3}{#4}
  \end{floatrow}
}
\newcommand{\figNtable}[4]{%
  \begin{floatrow}
    \CenterFloatBoxes
    \ffigbox{#1}{#2}
    \ttabbox{#3}{#4}
  \end{floatrow}
}

\newcommand{\Alu}[1]{%
\scalebox{.85}{
  \psset{unit=1mm}
  \begin{pspicture}(0,0)(12,6)
  \pspolygon(2,0)(10,0)(12,6)(0,6)
  \rput(6,3){#1}
  \end{pspicture}}%
}

\usepackage[ruled]{algorithm2e}

\SetAlFnt{\small}
\SetAlCapFnt{\small}
\SetAlCapNameFnt{\small}
\SetAlCapHSkip{0pt}
\IncMargin{-\parindent}

\setcopyright{none}

\begin{document}

\markboth{V. Jatala et al.}{Scratchpad Sharing in GPUs}

\title{Scratchpad Sharing in GPUs}
\author{Vishwesh Jatala
\affil{Indian Institute of Technology, Kanpur}
Jayvant Anantpur
\affil{Indian Institute of Science, Bangalore}
Amey Karkare
\affil{Indian Institute of Technology, Kanpur}}

\begin{abstract}

GPGPU applications exploit on-chip scratchpad memory available 
in the Graphics Processing Units (GPUs) to 
improve performance.  The amount of thread level parallelism (TLP)
present in the GPU is limited by the number of resident threads, which
in turn depends on the availability of scratchpad memory in its
streaming multiprocessor (SM).  Since the scratchpad memory is
allocated at thread block granularity, part of the memory may remain
unutilized.  In this paper, we propose architectural and compiler
optimizations to improve the scratchpad memory utilization.  Our
approach, called \emph{Scratchpad Sharing}, addresses scratchpad
under-utilization by launching additional thread blocks in each SM.
These thread blocks use unutilized scratchpad memory and also share 
scratchpad memory with other resident blocks. To improve the performance 
of scratchpad sharing, we propose \emph{Owner Warp First (OWF)} 
scheduling that schedules warps from the additional 
thread blocks effectively.  The performance of
this approach, however, is limited by the availability of the part of
scratchpad memory that is shared among thread blocks.

We propose compiler optimizations to improve the availability of
shared scratchpad memory.  We describe a scratchpad allocation scheme
that helps in allocating scratchpad variables such that shared
scratchpad is accessed for short duration.  We introduce a new
hardware instruction, {\em \Release}, that when executed, releases the
shared scratchpad memory.  Finally, we describe an analysis for
optimal placement of \Release\/ instructions such that shared
scratchpad memory is released as early as possible, but only after its
last use, along every execution path.

We implemented the hardware changes required for scratchpad sharing approach 
and the new instruction (\Release) using the GPGPU-Sim simulator, 
and implemented the compiler optimizations in
Ocelot framework.  We evaluated the effectiveness of our approach on
19 kernels from 3 benchmarks suites: CUDA-SDK, GPGPU-Sim, and Rodinia.
The kernels that under-utilize scratchpad memory show
an average improvement of
19\% and  maximum improvement of  92.17\% compared to  the baseline
approach, without affecting the performance of the kernels that do not
waste scratchpad memory.

\end{abstract}

%
%


\runningfoot{}

\begin{bottomstuff}
 
Vishwesh Jatala is supported by TCS Ph.D. fellowship. Jayvant Anantpur acknowledges the funding received from Google India Private Limited.

This article is extension of the paper ``Improving GPU Performance Through Resource Sharing", in Proceedings of the 25th ACM International Symposium on High-Performance Parallel and Distributed Computing (HPDC '16). The paper describes a resource sharing technique that makes architectural modifications to improve GPU performance. This work extends the paper by introducing compiler optimizations to leverage the resource sharing approach.

Author's addresses: Vishwesh Jatala, Department of Computer Science and Engineering,  Indian Institute of Technology, Kanpur; Jayvant Anantpur,
Supercomputer Education and Research Centre (SERC) , Indian Institute of Science, Bangalore; Amey Karkare, Department of Computer Science and Engineering,  Indian Institute of Technology, Kanpur; 
\end{bottomstuff}

\maketitle

\section{Introduction}
The throughput achieved by a GPU (Graphics Processing Unit) depends on
the amount of thread-level-parallelism (TLP) it utilizes. Therefore,
improving the TLP of GPUs has been the focus of many recent
studies~\cite{SharedMemMultiplexing, NMNL, Linearization,
  UnifiedOnChip}. The TLP present in a GPU is dependent on the number
of resident threads. A programmer interested in parallelizing an
application in GPU invokes a function, called {\em kernel}, with a
configuration consisting of number of thread blocks and number of
threads in each thread block. The maximum number of thread blocks, and
hence the number of threads, that can be launched in a Streaming
Multiprocessor (SM) depends on the number of available resources in
it.  If an SM has R resources and each thread block requires $R_{tb}$
resources, then $\left\lfloor{R}/{R_{tb}}\right\rfloor$ number of
thread blocks can be launched in each SM. Thus utilizing
$R_{tb}*\left\lfloor{R}/{R_{tb}}\right\rfloor$ units of resources
present in the SM; the remaining $R\text{ mod }R_{tb}$ resources are
wasted.
In this paper, we propose an approach, \emph{Scratchpad Sharing}, that
launches additional thread blocks in each SM.  These thread blocks
help in improving the TLP by utilizing the wasted scratchpad memory
and by sharing the scratchpad memory with the other resident thread
blocks. We further propose \emph{Owner Warp First (OWF)}, a warp
scheduling algorithm that improves performance by effectively
scheduling warps from the addition thread blocks.

In our experiments we observed that the performance of scratchpad
sharing depends on the availability of the scratchpad memory that is
shared between the thread blocks.  We have developed static analysis
that helps in allocating scratchpad variables into shared and unshared
scratchpad regions such that the shared scratchpad variables are
needed only for a short duration.  We modified the GPU architecture to
include a new hardware instruction (\Release) to release the acquired
shared scratchpad memory at run time.  When all the threads of a
thread block execute the \Release\/ instruction, the thread block
releases its shared scratchpad memory. We describe an algorithm to help
compiler in an optimal placement of the \Release\/ instruction in a
kernel such that the shared scratchpad can be released as early as
possible, without causing any conflicts among shared thread blocks.
These optimizations improve the availability of shared scratchpad
memory.

The main contributions of this paper are:
\begin{enumerate}
\item We describe an approach to launch more thread blocks by sharing
  the scratchpad memory. We further describe a warp scheduling
  algorithm that improves the performance of the GPU applications by
  effectively using warps from additional thread blocks.
\item We present a static analysis to layout scratchpad variables in
  order to minimize the shared scratchpad region. We introduce a
  hardware instruction, \Release, and an algorithm for optimal
  placement of \Release\/ in the user code to release the shared
  scratchpad region at the earliest.
\item We used the GPGPU-Sim~\cite{GPGPU-Sim} simulator and the
  Ocelot~\cite{Ocelot} compiler framework to implement and evaluate
  our proposed ideas. On several kernels from various benchmark
  suites, we achieved an average improvement of 19\% and a maximum
  improvement of 92.17\% over the baseline approach.
\end{enumerate}

The rest of the paper is organized as follows:
Section~\ref{sec:background} describes the background required for our
approach. Section~\ref{sec:scratchpadsharing} motivates the need for
scratchpad sharing and presents the details of the
approach. Owner Warp First scheduling is described in Section~\ref{sec:owf}. Section~\ref{sec:motivation} presents the need for compiler
optimizations. The optimizations themselves are discussed in
Section~\ref{sec:implementation}. Section~\ref{sec:analysis} analyzes the hardware requirements and the complexity of our approach. Section~\ref{sec:experimentanalysis}
shows the experimental results.  Section~\ref{sec:relatedwork}
discusses related work, and Section~\ref{sec:conclusion} concludes the
paper.

\section{Background} \label{sec:background}


A typical  NVIDIA GPU consists  of a set of  streaming multiprocessors
(SMs). Each  SM contains execution  units called stream  processors. A
programmer  parallelizes  an  application  on  GPU  by  specifying  an
execution configuration consisting of the  number of thread blocks and
the  number of  threads in  each thread  block. The  number of  thread
blocks that  are actually launched  in a  SM depends on  the resources
available in  the SM,  such as  the amount  of scratchpad  memory, the
number  of registers.   The  threads in a SM  are
grouped  into 32  threads,  called warps. All the threads in  a warp execute
the same instruction in SIMD manner.  GPU has  one  or more  warp
schedulers, which fetch a warp from  the pool of available warps based
on some warp  scheduling algorithm. When no warp can be issued in a 
cycle, the cycle is said to be a stall cycle.

\begin{table}[t]
\newcounter{magicrownumbers}
\newcommand\rownumber{\stepcounter{magicrownumbers}\arabic{magicrownumbers}}

\caption{Benchmark Applications for which the Number of Thread Blocks is Limited by Scratchpad Memory}
\vskip -3mm
\centering
\scalebox{0.75}{\renewcommand{\arraystretch}{1.05}
\begin{tabular}{r@{\quad}l@{\ }l@{\ }l@{\quad}r@{\quad}r@{\quad}r}
\hline\hline
&Benchmark &  Application & Kernel & \#Scratchpad & Scratchpad   & Block \\
&          &              &        & Variables &  Size (Bytes) & Size  \\
\hline
\multicolumn{7}{c}{{\bf Set-1: Shared scratchpad can be released before the end of the kernel}} \\ \hline
\rownumber. & RODINIA & backprop & bpnn\_layerforward\_CUDA & 2 & 9408 & 256\\
\rownumber. & CUDA-SDK & dct8x8\_1 (DCT1) & CUDAkernel2DCT & 1 & 2112 & 64\\
\rownumber. & CUDA-SDK & dct8x8\_2 (DCT2) & CUDAkernel2IDCT & 1 & 2112 & 64 \\
\rownumber. & CUDA-SDK & dct8x8\_3 (DCT3) & CUDAkernelShortDCT & 1 & 2176 & 128 \\
\rownumber. & CUDA-SDK & dct8x8\_4 (DCT4) & CUDAkernelShortIDCT & 1 & 2176 & 128 \\
\rownumber. & GPGPU-SIM & NQU & solve\_nqueen\_cuda\_kernel & 5 & 10496 & 64 \\
\rownumber. & RODINIA & srad\_v2\_1 (SRAD1) & srad\_cuda\_1 & 6 & 13824 & 576 \\
\rownumber. & RODINIA & srad\_v2\_2 (SRAD2) & srad\_cuda\_2 & 5 & 11520 & 576 \\
\hline
\multicolumn{7}{c}{{\bf Set-2: Shared scratchpad can not be released before the end of the kernel}} \\ \hline
\rownumber. & CUDA-SDK & FDTD3d & FiniteDifferencesKernel & 1 & 3840 & 128 \\
\rownumber. & RODINIA & heartwall & kernel & 8 & 11872 & 128 \\
\rownumber. & CUDA-SDK & histogram & histogram256Kernel & 1 & 9216 & 192 \\
\rownumber. & CUDA-SDK & marchingCubes (MC1) & generateTriangles & 2 & 9216 & 32 \\
\rownumber. & RODINIA & nw\_1 & needle\_cuda\_shared\_1 & 2 & 8452 & 32 \\
\rownumber. & RODINIA & nw\_2 & needle\_cuda\_shared\_2 & 2 & 8452 & 32 \\
\hline
\end{tabular}}
\label{table:set_1}
\vskip -3mm
\end{table}

NVIDIA provides a programming language CUDA~\cite{CUDA}, which can be
used to write an application to be parallelized on GPU.  The region of
a program which is to be parallelized is specified using a function
called kernel. The kernel is invoked with the configuration specifying
the number of thread blocks and number of threads as \linebreak {\tt
  <<<\#ThreadBlocks, \#Threads>>>}.  A variable can be allocated to 
  global memory by invoking {\tt
  cudamalloc()} function.  Similarly a variable can be allocated to
scratchpad memory by specifying {\tt \_\_shared\_\_} keyword inside a kernel
function. The latency of accessing a variable from global memory is 
400-800 cycles, whereas, latency of accessing from scratchpad memory 
is 20-30x lower than that of global memory~\cite{CUDA}.


\begin{figure}[t]
  	\includegraphics[scale=0.5]{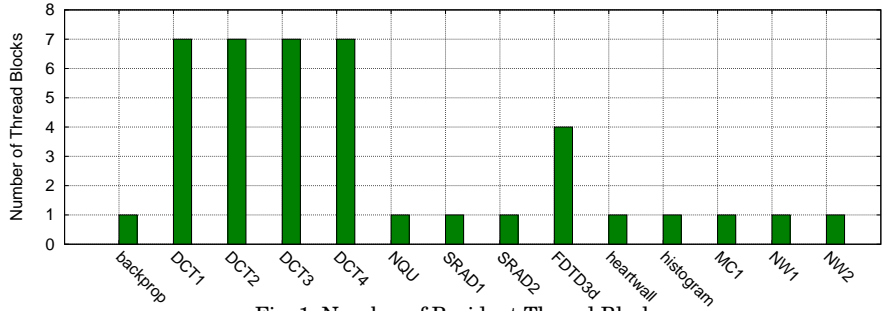}
	\vskip -3mm
    \caption{Number of Resident Thread Blocks}
    \label{fig:resident}
    \vskip -3mm          
\end{figure}

\section{Scratchpad Sharing} \label{sec:scratchpadsharing}

\begin{figure}[t]
{\includegraphics[scale=0.5]{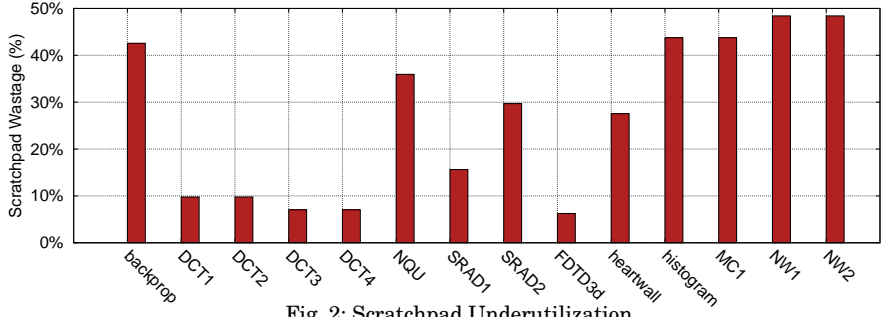}}
	\vskip -3mm
	\caption{Scratchpad Underutilization}
	\label{fig:wastage}
	\vskip -3mm          
\end{figure}

Scratchpad memory allocation at thread block level granularity causes
scratchpad underutilization. To understand the utilization of
scratchpad memory, we analyzed applications shown in
Table~\ref{table:set_1}.  Figure~\ref{fig:resident} shows the number
of thread blocks that are launched in each SM and
Figure~\ref{fig:wastage} shows the percentage of unutilized scratchpad
memory for the GPU configuration shown in Table~\ref{table:GPGPUArch}.

\begin{example} \label{ex:motivation}
Consider the application \emph{backprop} in
Table~\ref{table:set_1}. It requires 9408 bytes of scratchpad memory
to launch a thread block in the SM. According to the GPU configuration
shown in Table~\ref{table:GPGPUArch}, each SM has 16K bytes of
scratchpad memory. Hence only 1 thread block can be launched in the
SM, this utilizes 9408 bytes of scratchpad memory. The remaining 6976
bytes of scratchpad memory remains unutilized. We can observe the
similar behavior for other applications as well. Hence scratchpad
allocation at thread block level granularity not only has lower number
of resident thread blocks but also has scratchpad memory
underutilization.\hfill\qed
\end{example}

To address this problem, we propose, \emph{Scratchpad Sharing}, that
increases the number of resident thread blocks in each SM.  These
thread blocks use the unutilized scratchpad memory as well as share
scratchpad with other resident thread blocks.  This not only reduces 
scratchpad underutilization but also increases TLP on the SM.

\begin{example} To  improve the performance of \emph{backprop}, we launch two
  thread blocks (say $\TB_0$ and $\TB_1$), which share the scratchpad
  memory.  Instead of allocating 9408 bytes of scratchpad memory to
  each of $\TB_0$ and $\TB_1$, the scratchpad sharing approach
  allocates total 16K bytes of memory together for $\TB_0$ and
  $\TB_1$. In this case, 6976 bytes (the unutilized amount in Example~\ref{ex:motivation}) of scratchpad memory is
  allocated to each thread block independently ({\em unshared}
  scratchpad), while the remaining 2432 bytes of memory ({\em
    shared} scratchpad) is allocated to the thread block which
  requires it first.  For example, if $\TB_1$ accesses the shared
  scratchpad memory first, it is allocated all of the shared portion.
  $\TB_0$ can continue its execution till it requires shared
  scratchpad memory, at which point it waits. $\TB_0$ resumes its
  execution once $\TB_1$ finishes or releases the shared scratchpad.
  Thus, $\TB_0$ can help in hiding the long memory latencies of
  $\TB_1$,
  thereby
  improving the run-time of the application.\hfill\qed
\end{example}

\begin{figure}[t]
  \figNtable{\scalebox{.79}{\small
  \psset{unit=1mm}
  \psset{linewidth=.3mm}
  \begin{pspicture}(0,-12)(75,40)
    \putnode{t0}{origin}{8}{25}{
      $\left(\begin{tabular}{@{}c@{}}
        ThId,\\ SMemLoc
    \end{tabular}\right)$}
    
    \putnode{t1}{t0}{33}{0}{\psdiabox[framesep=-3]{
        \rule[-6mm]{0cm}{12mm}\begin{tabular}{c}
          ThId $\in$ \\ Unshared \\ ThreadBlock?
        \end{tabular}
    }}
    
    \putnode{t2}{t1}{0}{-20}{\psdiabox[framesep=-3]{
        \rule[-4.5mm]{0cm}{10mm}\begin{tabular}{c}
          SMemLoc \\ $\leq R_{tb}t$?
        \end{tabular}
    }}

    \putnode{t3}{t2}{0}{-20}{\psdiabox[framesep=-3]{
        \rule[-4.5mm]{0cm}{10mm}\begin{tabular}{c}
          Acquired \\ Lock?
        \end{tabular}
    }}
    
    \putnode{t4}{t2}{27}{0}{
      \begin{tabular}[b]{|c|}\hline
        \rule{3mm}{0mm}\\\hline
        \\\hline
        \\\hline
        \\\hline
        \\\hline
      \end{tabular}\rotateleft{Scratchpad}
    }
    
    \putnode{a4}{t4}{0}{-15}{\Alu{ALU}}\ncline{->}{t4}{a4}

    \ncline{->}{t0}{t1}
    \ncline{->}{t1}{t2}\Aput[.2]{No}
    \ncline{->}{t2}{t4}\Aput[.2]{Yes}
    \ncline{->}{t2}{t3}\Bput[.2]{No (Shared Loc)}
    \ncangle[angleA=180,angleB=-90]{->}{t3}{t0}\Bput[.2]{No (Retry)}
    \ncangle[angleA=0,angleB=90]{->}{t1}{t4}\Bput[.2]{Yes}
    \ncangle[angleA=0,angleB=180,offsetB=-1.5]{->}{t3}{t4}\Aput[.2]{Yes}

    \putnode{ca}{t0}{0}{6}{(a)}
    \putnode{cb}{t1}{-9}{6}{(b)}
    \putnode{cc}{t2}{-7}{6}{(c)}
    \putnode{cd}{t4}{-8}{9}{(d)}
    \putnode{ce}{t3}{-8}{-5}{(e)}
\end{pspicture}} \vskip 3mm}
            {\caption{Scratchpad Access Mechanism\label{fig:spaccess}}}
            {\scalebox{0.86}{\renewcommand{\arraystretch}{.99}
                \begin{tabular}{@{}l|l@{\ }}
                  \hline\hline
                  Resource & Configuration \\
                  \hline
                  Number of Clusters & 14 \\
                  Number of Cores/Cluster & 1 \\
                  Core Clock	& 732 MHz \\ 
                  Scratchpad Memory/Core & 16KB \\
                  Number of Registers/Core \space \space \space & 65536  \\ 
                  Max Number of TBs/Core & 16 \\
                  Max Number of Threads/Core & 3072  \\
                  Warp Scheduling & LRR \\
                  Number of Schedulers/Core & 4 \\
                  L1-Cache/Core & 16KB \\
                  L2-Cache & 1.5MB \\
                  DRAM Scheduler & FR-FCFS  \\ \hline
              \end{tabular}}}
            {\caption{GPGPU-Sim Architecture}\label{table:GPGPUArch}}

  \vskip -3mm
\end{figure}

To generalize our idea, consider a GPU that has $R$ units of 
scratchpad memory per SM, and
each thread block requires $R_{tb}$ units of scratchpad memory to
complete its execution. Consider a pair $\TB_0$ and $\TB_1$ of shared
thread blocks. Instead of allocating $R_{tb}$ units of memory to each
of $\TB_0$ and $\TB_1$, we allocate $t\times R_{tb}$ ($0 < t < 1$)
units of scratchpad memory to each of them independently.  This is
called \emph{unshared} scratchpad.  We further allocate $(1-t)\times
R_{tb}$ units of scratchpad memory to the pair as \emph{shared}
scratchpad. Thus, a total of $(1+t)\times R_{tb}$ units of scratchpad
memory is allocated for both.  $\TB_0$ and $\TB_1$ can access shared
scratchpad memory only after acquiring an exclusive lock, in an FCFS
manner, to prevent concurrent accesses. Once a shared thread block
(say $\TB_0$) acquires the lock for shared scratchpad memory, it
retains the lock till the end of its execution. The other thread
block ($\TB_1$) can continue to make progress until it requires to
access shared scratchpad memory, at which point it waits until $\TB_0$
releases the shared scratchpad.

The naive scratchpad sharing mechanism, where each thread block shares
scratchpad memory with another resident thread block, may not give
benefit over default (unshared) approach. we also need to guarantee
that in sharing approach, the number of active thread blocks (not
waiting for shared scratchpad) is no less than the number of thread
blocks in default approach.
\begin{example}
  Consider the application DCT3 that requires 2176 bytes of scratchpad
  memory per thread block. For the given GPU configuration
  (Table~\ref{table:GPGPUArch}), 7 thread blocks can be launched in
  default mode. With scratchpad sharing, it is possible to launch 12
  thread blocks (for a certain value of $t$). Suppose we create 6 pairs of thread blocks where the
  blocks in each pair share scratchpad. Then, in the worst case, all 12
  blocks may request access to the shared portion of scratchpad. This
  will cause 6 blocks to go in waiting, while only the remaining 6
  will make progress. If the shared region is sufficiently large, the
  application will perform worse with scratchpad sharing.

  To make sure at least 7 thread blocks make progress, our approach
  creates only 5 pairs of thread blocks that share scratchpad memory, 
  the remaining 2 thread blocks are not involved in sharing. 
  Thus, at most 5 blocks   can be waiting during execution.\hfill\qed
\end{example}

In our approach, the thread blocks that share the scratchpad memory
are referred to as {\em shared thread blocks}, the rest are referred
to as {\em unshared thread blocks}. The computation of number of
shared and unshared thread blocks is described in detail
in~\cite{resourcesharing}.

To implement our approach, we modify the existing scratchpad access
mechanism provided by \citeN{GPGPUSIM} simulator. Figure~\ref{fig:spaccess} shows the
scratchpad access mechanism that supports scratchpad sharing. When a
thread (Thread Id: $ThId$) needs to access a scratchpad location
($SMemLoc$), we need to check if it is from an unshared thread
block. If it belongs to an unshared thread block, it can access the
location directly from scratchpad memory (Figure~\ref{fig:spaccess} Step
(b)). Otherwise, we need to make another check if it accesses unshared
scratchpad location (Step (c)). The thread accesses unshared
scratchpad location if $SMemLoc < R_{tb}t $ because we allocate
$R_{tb}t$ units of scratchpad memory to each of the shared thread
blocks. Otherwise, we treat the location as shared scratchpad
location. A thread can access unshared scratchpad location directly,
however it can access the shared scratchpad location only after
acquiring the exclusive lock as shown in Step (e). Otherwise, it retries
the access in the next cycle\footnote{The details of required additional storage
units are discussed in Section~\ref{sec:analysis}.}.

\section{Owner Warp First (OWF) Scheduling}
\label{sec:owf}

In our approach, each SM contains various types of thread blocks such as, (1) unshared thread blocks, which do not share scratchpad memory with any other thread block, (2) shared thread blocks that own the shared scratchpad memory (\emph{Owner thread blocks}) by having exclusive lock, and (3) shared thread blocks that do not own the shared scratchpad memory (\emph{Non-owner thread blocks}). We refer the to the warps in these thread blocks as `Unshared Warps', `Owner warps', and `Non-owner Warps' respectively. When an owner thread block finishes its execution, it transfers its ownership to its corresponding non-owner thread block, and the new thread block that will be launched becomes the non-owner thread block.

\begin{figure}[t]
  \includegraphics[scale=0.4]{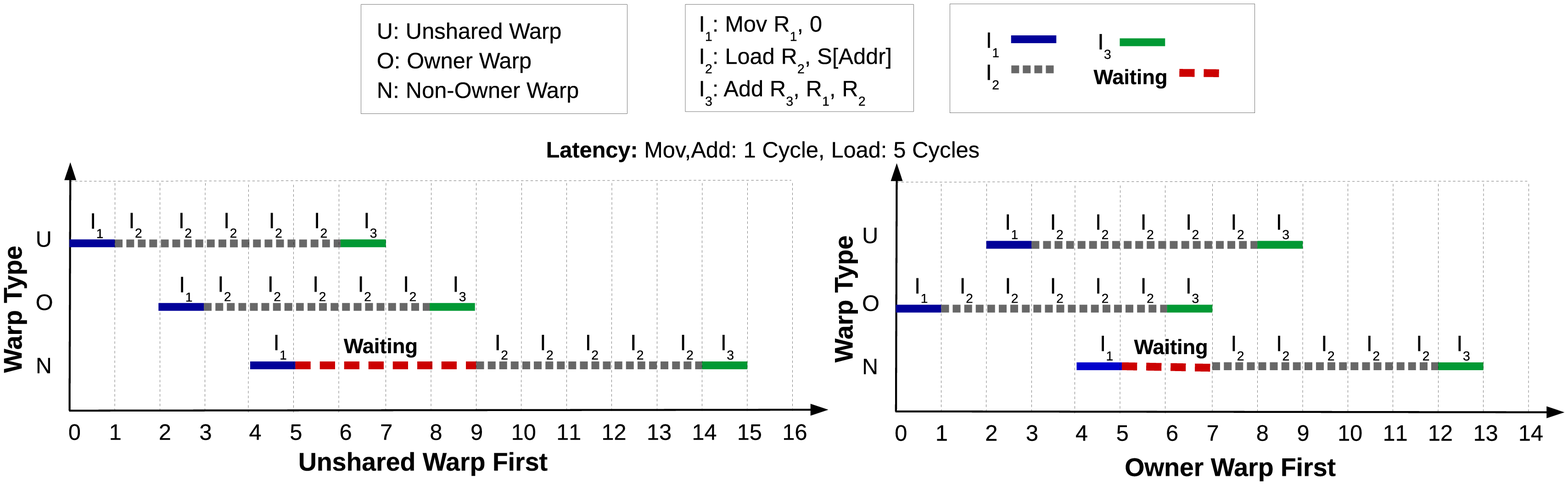} \vskip -4mm
  \caption{Warp Scheduling}\label{fig:OWF}\vskip -2mm
\end{figure}

Scheduling these warps in the SM plays an important role in improving the performance of applications. Hence we propose an optimization, \emph{Owner Warp First (OWF)}, that schedules the warps in the following order: (1) Owner warps, (2) Unshared warps, and (3) Non-owner warps. Giving the highest priority to owner warps helps them in finishing sooner so that the dependent non-owner warps can resume their execution, which can help in hiding long execution latencies. Figure~\ref{fig:OWF} shows the benefit of giving the first priority to owner warp when compared to non-owner warp. Consider an SM that has three warps: Unshared (U), Owner (O), and Non-owner (N) warps. Assume that they need to execute 3 instructions $I_1, I_2$, and $I_3$ as shown in the figure; the latency of \emph{Add} and \emph{Mov} instruction is 1 cycle, and the latency of \emph{Load} instruction is 5 Cycles. Also, assume that instruction $I_2$ uses a shared scratchpad memory location. If unshared warp is given highest priority (shown as Unshared Warp First in the figure), then it can issue $I_1$ in the 1st cycle and issue $I_2$ in the 2nd cycle. However, it can not issue $I_3$ in the 3rd cycle since $I_3$ is dependent on $I_2$ for register $R_2$, and $I_2$ takes five cycles to complete its execution. However, the owner warp can execute $I_1$ in the 3rd cycle. Similarly, it can issue $I_2$ in the fourth cycle. The non-owner with least priority can start issuing $I_1$ in the 5th cycle, however, it can not issue $I_2$ in the 6th cycle since $I_2$ uses a shared scratchpad memory location, and it can access the shared location only after the owner thread block releases the lock, hence it waits until the owner warp finishes execution. Once the owner warp finishes the execution of $I_2$ and $I_3$ in the 8th and 9th cycles respectively, the non-owner resumes the execution of $I_2$ in 10th cycle, and it can subsequently  finish in 15 cycles.

If owner warp is given first priority compared to unshared warp, it can issue $I_1$ and $I_2$ in its 1st and 2nd cycles respectively. Similarly, the unshared warp, with second priority, can issue $I_1$ and $I_2$ in 3rd and 4th cycles. The non-owner warp with least priority can issue $I_1$ in 5th cycle, and it waits for owner warp to release the shared scratchpad memory. Once the owner warp completes the execution of $I_2$ and $I_3$in 7th and 8th cycles, the non-owner can resume the execution by overlapping the execution of $I_2$ in the 8th cycle with unshared warp. Finally, the unshared warp and non-owner warp can finish their execution in 9th and 13th cycle respectively. Thus improving the overall performance.

\subsection{Effect of Barriers}
\newcommand{\barrier}{{\tt \_syncthreads()}} 

Our approach to scratchpad sharing does not require any special
handling of barriers (\barrier).  Recall that in our approach,
the scratchpad is shared at the thread block granularity.  An
owner thread block gets the lock on the shared scratchpad and
releases only after it completes its execution.  The warps from
the non-owner thread block wait for its owner thread block for
shared scratchpad memory.  However, the warps from the owner
thread block never wait for the warps of the non-owner thread
block since they already have lock.  In the presence of
\barrier\ instruction only the warps within the same thread
blocks wait for other warps, and they make progress after all the
warps of the thread block arrive at the barrier~\cite{CUDA}.
Hence a circular wait is not possible even in the presence of
both barriers as well as shared scratchpad locks, thus avoiding
deadlocks.

\subsection{Scratchpad Sharing on Multiple SMs}
\citeN{GPGPUSIM} uses a round robin scheduling algorithm for
scheduling thread blocks on multiple SMs\footnote{Note that this
  is different from NVIDIA GPU that is believed to use a
  FIFO policy~\cite{preemptive}.}. Consider a GPU that has $p$
SMs. Assume that a kernel needs to launch $N$ thread blocks
($B_1, B_2, \ldots, B_{N}$) in the GPU. Further, assume that
default scratchpad allocation mechanism can launch $m$ thread
blocks in each SM. Thus, the $i^{th}$ SM initially has thread
blocks with ids $B_i, B_{p+i}, B_{2p+i}, \ldots B_{(m-1)p+i}$ in
it.  Whenever a SM finishes the execution of a thread block, a
new thread block is launched in it, until all the $N$ blocks are
finished.

Our proposed scratchpad sharing mechanism does not modify the
thread block scheduling mechanism. Assuming our approach launches
$n$ blocks ($n \geq m$), the $i^{th}$ SM gets thread blocks with
ids $B_i, B_{p+i}, B_{2p+i}, \ldots, B_{(n-1)p+i}$. The
additional $n-m$ thread blocks (i.e., $B_{mp+i}, B_{(m+1)p+i},
\ldots$) share scratchpad memory with blocks $B_i, B_{p+i},
\ldots$ respectively. The remaining thread blocks remain in the
unsharing mode. Note that the thread blocks that share scratchpad
memory are always part of the the same SM. Whenever a new thread
block is launched in place of an old block that has
finished its execution, the new block gets the same
scratchpad sharing status (shared/unshared) as the old one.

Since thread blocks in an SM
can complete their execution in any order~\cite{CUDA} (no
priority among thread blocks), and the scratchpad memory is
shared only among the thread blocks within a SM, priority
inversion problem does not arise with the scratchpad sharing
approach.


\newcommand{\fgentry}{{\sf Entry}}
\newcommand{\fgexit}{{\sf Exit}}
\newcommand{\pt}{\ensuremath{\pi}}

\section{The Need for Compiler Optimizations} \label{sec:motivation}

\begin{figure}[t]
    \twofigure{\includegraphics[scale=0.37]{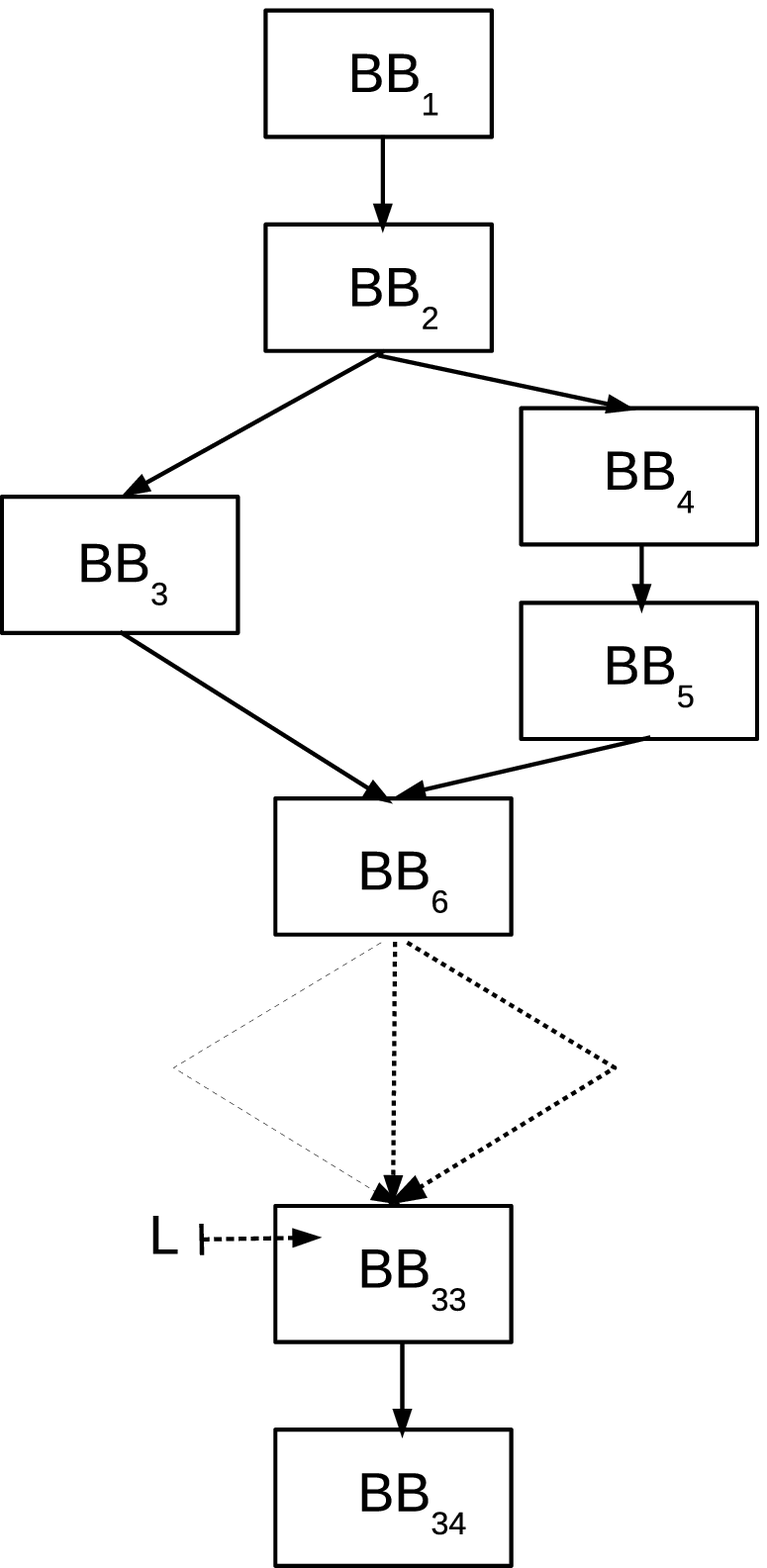}}
           {\caption{Release of Shared Scratchpad} \label{fig:release}}
           {\includegraphics[scale=0.37]{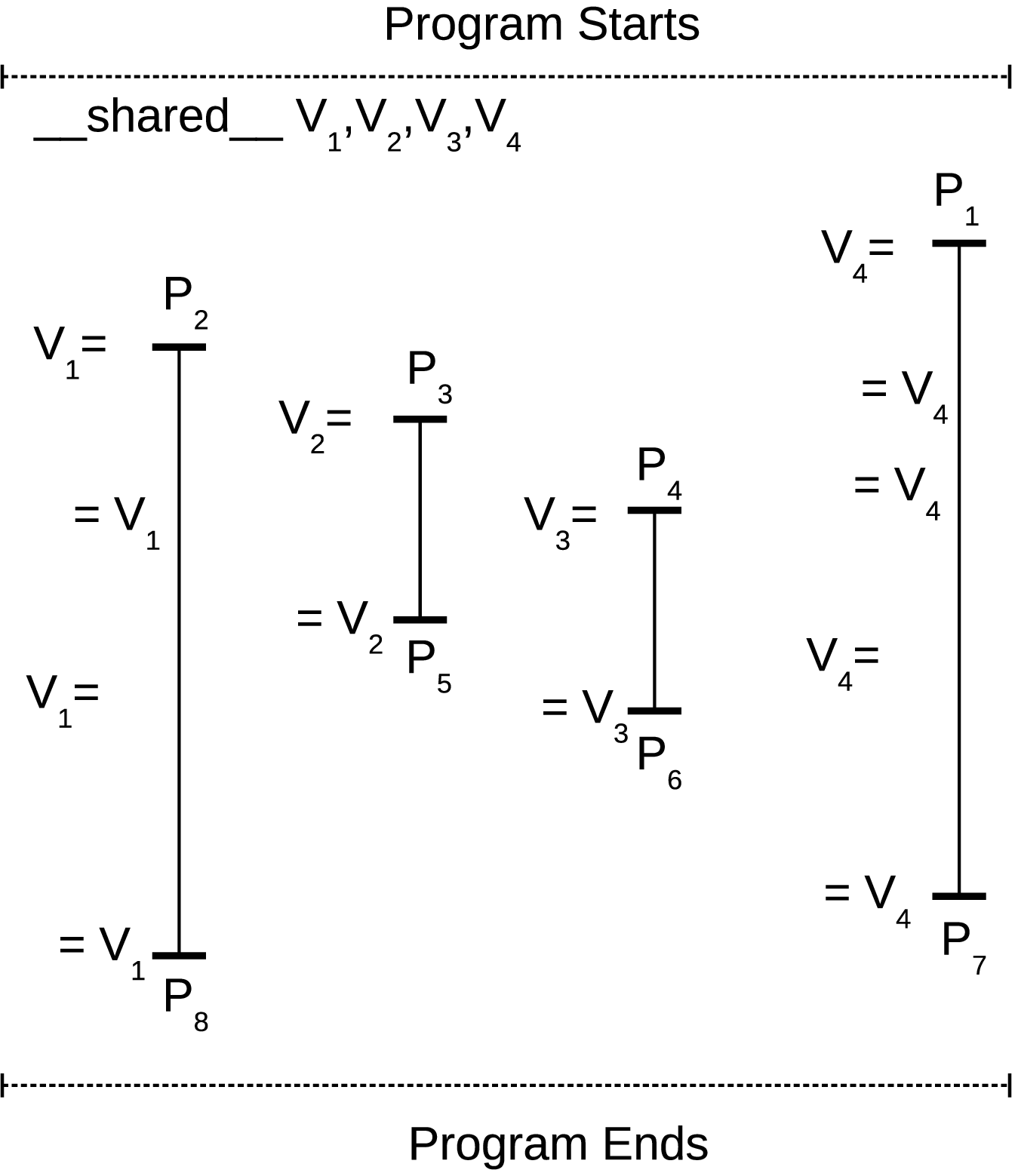}}
           {\caption{Access Ranges of Scratchpad Variables}\label{fig:intervals}}
\end{figure}

In scratchpad sharing, when two thread blocks (say, $\TB_0$ and $\TB_1$) are launched in
shared mode, one of them accesses the shared scratchpad region at a
time. As soon as one thread block, say $\TB_0$, starts accessing the
shared scratchpad region, the other thread block, $\TB_1$, can not access the 
shared scratchpad region and hence may have wait until $\TB_0$ finishes execution.

\begin{example}
Consider the CFG in Figure~\ref{fig:release}, which is obtained for
\emph{SRAD1} benchmark application (Table~\ref{table:set_1}).  In the
figure, the program point marked {\bf L} corresponds to the last
access to the shared scratchpad.  Without compiler assistance, the
shared scratchpad region can be released only at the end of the last
basic block (\fgexit\/ node of CFG) even though it is never accessed
after L.\hfill\qed
\end{example}
To promote the release of shared scratchpad region before the end of
kernel execution, we introduce a new hardware instruction (PTX
instruction) called \Release.  Our proposed compiler optimization can
place the \Release\/ instructions in a kernel such that shared
scratchpad memory is released as early as possible by each thread.

\begin{example} \label{ex2}
Consider the scenario in Figure~\ref{fig:intervals}, where a kernel
function declares four equal sized scratchpad variables $V_1$ to
$V_4$. The figure also shows the regions of the kernel within which
different variables are accessed.  If $V_1$
and $V_4$ are allocated into shared scratchpad region, then the shared
scratchpad region is accessed from program point $P_1$ to program
point $P_8$. However, when $V_2$ and $V_3$ are allocated to shared
scratchpad region, the shared region is accessed for a shorter
duration, i.e., from program point $P_3$ to program point
$P_6$.\hfill\qed
\end{example}
Note that the choice of allocation of scratchpad variables into
shared and unshared scratchpad regions does not affect the
correctness of the program but can affect the availability of the
shared scratchpad region, and hence the effectiveness of sharing.

\section{Compiler Optimizations} \label{sec:implementation}
In this section we describe a compile time memory allocation
scheme and an analysis to optimally place \Release\ instructions.
The memory allocation scheme allocates scratchpad variables into
shared and unshared region such that shared scratchpad variables
are accessed only for a small duration during the run-time.  In
the presence of loops where the number of iterations of loops are
not computable at compile time, it is not possible to statically
bound the number of instructions executed at run-time. Hence we
need to use approximate loop bounds.  Any approximation is fine
since, as noted earlier, it only affects the effectiveness of
sharing, but not the correctness\footnote{Profiling and user
  annotations can help in finding better approximations for the
  loop bounds. However we have not used these in our current
  implementation.}.

To simplify the description of the required analyses, we make the
following assumptions:
\begin{itemize}
\item  The control flow  graph  (CFG) for  a function  (kernel)  has a  unique
  \fgentry\/ and a unique \fgexit\/ node.
\item There are no {\em critical} edges in the CFG.  A critical
  edge is an edge whose  source node has more than one successor
  and the destination node has more than one predecessor.
\end{itemize}
These assumptions are not restrictive as any control flow graph
can be converted to the desired form using a preprocessing step
involving simple graph transformations: adding a source node,
adding a sink node, and adding a node to split an
edge~\cite{Khedker.DFA,globalDFA,Muchnick}.

\subsection{Minimizing Shared Scratchpad Region}\label{sec:minregion}

Consider   a     GPU      that     uses      scratchpad     sharing
approach such  that two thread  blocks involved
in sharing can share a fraction  $f < 1$ of scratchpad memory.  Assume
that each SM in the GPU has $M$ bytes of scratchpad memory, the kernel
that is to  be launched into the SM has  $N$ scratchpad variables, and
each thread block of the  kernel requires $M_{tb}$ bytes of scratchpad
memory.   We allocate  a  subset  $S$ of  scratchpad
variables  into  shared  scratchpad   region  such that:
\begin{enumerate*}
\item  The  total size of  the scratchpad variables  in the set  $S$ is
  equal to  the size  of shared  scratchpad   ($f \times  M_{tb}$),
  and \label{cond:size}
\item The region of access for variables in $S$ is minimal in terms of
  the number of instructions.\label{cond:durn}
\end{enumerate*}

To compute the region of access  for $S$, we define {\em access range}
for a variable as follows:

\begin{definition}{\bf Access Range of a Variable:}
A program point $\pt$  is in the {\em access range}  of a variable $v$
if both  the following conditions hold:  
\begin{enumerate*}
\item There is an access (definition or use) of $v$ on some path
  from \fgentry\/ to \pt\ {\em and}
\item There is an access of $v$ on some path from \pt\ to
  \fgexit.
\end{enumerate*}
\end{definition}

Intuitively, the access range of  a variable covers every program
point  between  the first  access  and  the  last access  of  the
variable in an  execution path. The access  range for a
variable can contain disjoint regions due to branches in the flow
graph.

\begin{definition}{\bf Access Range of a Set of Variables:}
  A program point $\pt$ is in the  {\em access range} of a set of
  variable $S$ if both the following
  conditions hold:
\begin{enumerate*}
\item There  is an access  to a variable $v \in S$ on
  a path from \fgentry\/ to \pt\ {\em and}
\item There is an access to a variable $v' \in S$ on a path from
  \pt\ to \fgexit.  
\end{enumerate*}
\end{definition}

\begin{figure}[t]
    \figNtable{\includegraphics[scale=0.33]{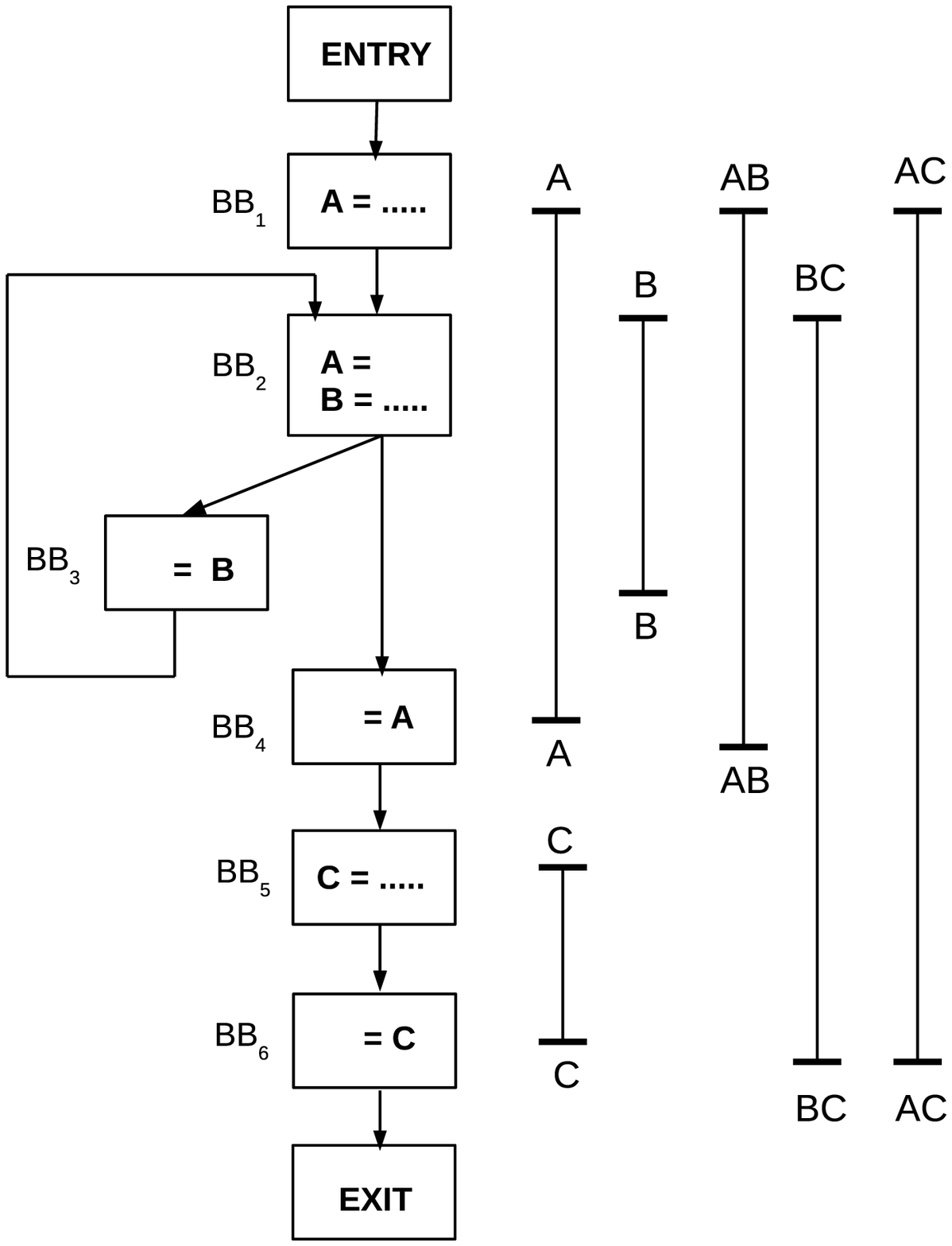} \vskip 2mm}
    {\vskip -4mm \caption{Access Ranges of Variables}\label{fig:accessrange}}
    {\scalebox{0.9}{\renewcommand{\arraystretch}{1.2}
        \begin{tabular}{@{}r@{\ }||@{\ }c@{\ }|@{\ }c@{\ }|@{\ }c@{\ }|@{\ }c@{\ }|@{\ }c@{\ }|@{\ }c||@{\ }c@{\ }|@{\ }c@{\ }|@{\ }c@{\ }|@{\ }c@{\ }|@{\ }c@{\ }|@{\ }c@{}}
          \hline 
          & \multicolumn{6}{@{}c@{}||}{For Variables} 
          & \multicolumn{6}{@{}c@{}}{For Sets of Variables}\\ \cline{2-13}
          & \multicolumn{3}{@{}c@{}|}{IN} 
          & \multicolumn{3}{@{}c@{}||}{OUT} 
          & \multicolumn{3}{@{}c@{}|}{IN} 
          & \multicolumn{3}{@{}c@{}}{OUT} \\ \cline{2-13}
          Block& A&B&C& A&B&C& AB& BC& CA& AB& BC& CA \\
          \hline\hline
           \fgentry & f&f&f& f&f&f&  f&  f&  f&  f&  f&  f \\
           BB1       & f&f&f& t&f&f&  f&  f&  f&  t&  f&  t \\
           BB2       & t&t&f& t&t&f&  t&  t&  t&  t&  t&  t \\
           BB3       & t&t&f& t&t&f&  t&  t&  t&  t&  t&  t \\
           BB4       & t&f&f& f&f&f&  t&  t&  t&  f&  t&  t \\
           BB5       & f&f&f& f&f&t&  f&  t&  t&  f&  t&  t \\
           BB6       & f&f&t& f&f&f&  f&  t&  t&  f&  f&  f \\
                    \fgexit  & f&f&f& f&f&f&  f&  f&  f&  f&  f&  f \\
                    \hline
        \end{tabular}}}
    {\caption{Access Ranges for Scratchpad Variables and Sets.\\ {\bf
          t} denotes true, {\bf f} denotes false. Sets of variables
        are written as concatenation of variables. For example,
        AB denotes \{A, B\}.\label{table:acc-range}}}
        \vskip -2mm
  \end{figure}

\begin{example}\label{ex:acc-range}
  Consider a kernel whose CFG is shown in
  Figure~\ref{fig:accessrange}. The kernel uses 3 scratchpad
  variables A, B and C.  Variable A is accessed in the region
  from basic block $BB_1$ to basic block $BB_4$. The start of
  basic block $BB_2$ is considered in access range of A because
  there is a path from \fgentry\/ to start of $BB_2$ that
  contains an access of A (definition in $BB_1$) and there is a
  path from the start of $BB_2$ to \fgexit\ that contains the
  access of A (use in $BB_4$).

  Consider the set $S =$ \{B, C\}.  Basic block $BB_4$ is in
  access range of $S$ because there is a path from \fgentry\/ to
  $BB_4$ containing the access of B (definition in $BB_2$), and
  there is a path from $BB_4$ to \fgexit\/ containing the access
  of the C (use in $BB_6$).\hfill\qed
\end{example}

To compute the access ranges for a program, we need a forward analysis
to  find the first access  of  the  scratchpad variables,  and  a
backward analysis  to find  the last access of  the scratchpad  variables. We
define these analyses formally using the following notations:

\newcommand{\indef}[2]{\ensuremath{{\sf PreIN}(#1, #2)}}
\newcommand{\outdef}[2]{\ensuremath{{\sf PreOUT}(#1, #2)}}
\newcommand{\inuse}[2]{\ensuremath{{\sf PostIN}(#1, #2)}}
\newcommand{\outuse}[2]{\ensuremath{{\sf PostOUT}(#1, #2)}}
\newcommand{\inacc}[2]{\ensuremath{{\sf AccIN}(#1, #2)}}
\newcommand{\outacc}[2]{\ensuremath{{\sf AccOUT}(#1, #2)}}
\newcommand{\IN}[1]{\ensuremath{{\sf IN}(#1)}}
\newcommand{\OUT}[1]{\ensuremath{{\sf OUT}(#1)}}
\newcommand{\Pred}[1]{\ensuremath{{\sf PRED}(#1)}}
\newcommand{\Succ}[1]{\ensuremath{{\sf SUCC}(#1)}}
\newcommand{\ininst}[2]{\ensuremath{{\sf InstIN}(#1, #2)}}
\newcommand{\outinst}[2]{\ensuremath{{\sf InstOUT}(#1, #2)}}

\begin{itemize}
\item \IN{BB} denotes the program point before the first statement of
  the basic block $BB$. \OUT{BB}  denotes the program point after the
  last statement of $BB$.
\item  \Pred{BB}  denotes  the  set of  predecessors,  and
  \Succ{BB} denotes the set of successors of $BB$. 

\item \indef{v}{BB} is true if there is an access to variable $v$  
before \IN{BB}.  \outdef{v}{BB} is  true if  there is an access  to the variable $v$ before \OUT{BB}.
\item \inuse{v}{BB}  is true if there  is an access to  variable $v$ after
  \IN{BB}. \outuse{v}{BB} is  true if there is a access to variable $v$
  after \OUT{BB}.

\item \inacc{S}{BB} is true if \IN{BB} is  in access range of a set of
  scratchpad variables $S$.  \outacc{S}{BB} is true if  \OUT{BB} is in
  access range of a set of scratchpad variables $S$.

\end{itemize}
The data flow equations to compute the information
are\footnote{The analysis can be extended easily to compute
  information at any point inside a basic block. We ignore it for
  brevity.}:
\[\small\begin{array}{r@{\ }c@{\ }l@{\quad}r@{\ }c@{\ }l}
  \outdef{v}{BB} &=& \left\{\renewcommand{\arraystretch}{1.1}
  \begin{array}{@{}l}
    \mbox{true, if $BB$ has an access of $v$}\\
    \indef{v}{BB},\  \mbox{otherwise}
  \end{array}\right. &
  \indef{v}{BB} &=& \left\{\renewcommand{\arraystretch}{1.1}
  \begin{array}{@{}l}
    \mbox{false, if $BB$ is \fgentry\ block}\\
    \bigvee\limits_{BP \in \Pred{BB}}  \hspace*{-5mm}
    \outdef{v}{BP},\ \mbox{otherwise}
  \end{array}\right.
\end{array}\]
\[\small\begin{array}{r@{\ }c@{\ }l@{\quad}r@{\ }c@{\ }l}
  \inuse{v}{BB} &=& \left\{\renewcommand{\arraystretch}{1.1}
  \begin{array}{@{}l}
    \mbox{true, if $BB$ has an access of $v$}\\
    \outuse{v}{BB},\  \mbox{otherwise}
  \end{array}\right. &
  \outuse{v}{BB} &=& \left\{\renewcommand{\arraystretch}{1.1}
  \begin{array}{@{}l}
    \mbox{false, if $BB$ is \fgexit\ block}\\
    \bigvee\limits_{BS \in \Succ{BB}} \hspace*{-5mm}
    \inuse{v}{BS},\ \mbox{otherwise}
  \end{array}\right.
\end{array}\]
We decide whether the access range of a set of scratchpad variables $S$ includes the points \IN{BB} and \OUT{BB} as:
\[\renewcommand{\arraystretch}{1.2}
\small\begin{array}{@{}r@{}c@{}l}
\inacc{S}{BB} &=& (\bigvee\limits_{v \in S} \indef{v}{BB}) \bigwedge (\bigvee\limits_{v \in S} \inuse{v}{BB}) \\
\outacc{S}{BB} &=& (\bigvee\limits_{v \in S} \outdef{v}{BB}) \bigwedge (\bigvee\limits_{v \in S} \outuse{v}{BB}) 
\end{array}\]

\begin{example}\label{ex:acc-range-2}
Table~\ref{table:acc-range} shows the program points in the access ranges of scratchpad variables for CFG of Figure~\ref{fig:accessrange}. The table also shows the program points in the access ranges of sets of two scratchpad variables each.  \hfill\qed
\end{example}

Let $SV$ denote the set of all scratchpad variables. For every subset $S$ of $SV$ having a total size equal to the size of shared scratchpad memory, our analysis counts the total number of instructions in the access range of S. Finally the subset that has the minimum count is selected for allocation in the shared scratchpad memory. 

\begin{example}\label{ex:opt-alloc}
Consider  once again  the  CFG  in Figure~\ref{fig:accessrange}.   For
simplicity,  assume that  all  the variables  have  equal sizes,  
and each  basic block contains the same number of instructions. 
Consider a scratchpad  sharing approach that can allocate  only two of
the variables into  the shared scratchpad region.  From the  CFG, and from Table~\ref{table:acc-range}, it is
clear that when  A and B are allocated into  shared scratchpad memory,
the shared region  is smaller,  compared to when
either \{B,C\} or \{A,C\} are allocated in the shared region.\hfill\qed
\end{example}

\newcommand{\relcnt}{{\bf count}}

\subsection{Implementation of \Release\ Instruction} \label{sec:release}
In scratchpad sharing approach, a shared thread block acquires a lock
before accessing shared scratchpad region and unlocks it only after
finishing its execution. This causes a delay in releasing the shared
scratchpad because the thread block holds the scratchpad memory till
the end of its execution, even though it has finished accessing shared
region.

\begin{figure}[t]
  \twofigure
      {
\begin{lstlisting}[escapechar=@]
void ReleaseSSP() 
{
    static int @\relcnt@=0; 
    ++@\relcnt@;
    if(@\relcnt@==ACTIVE_THREADS){@\label{done}@
        @\relcnt@=0;
        @{\sc UnlockSharedRegion}@();
    }
}
\end{lstlisting}}
      {\caption{Pseudocode of \Release\ Instruction}\label{clear}}
      {\includegraphics[scale=0.36]{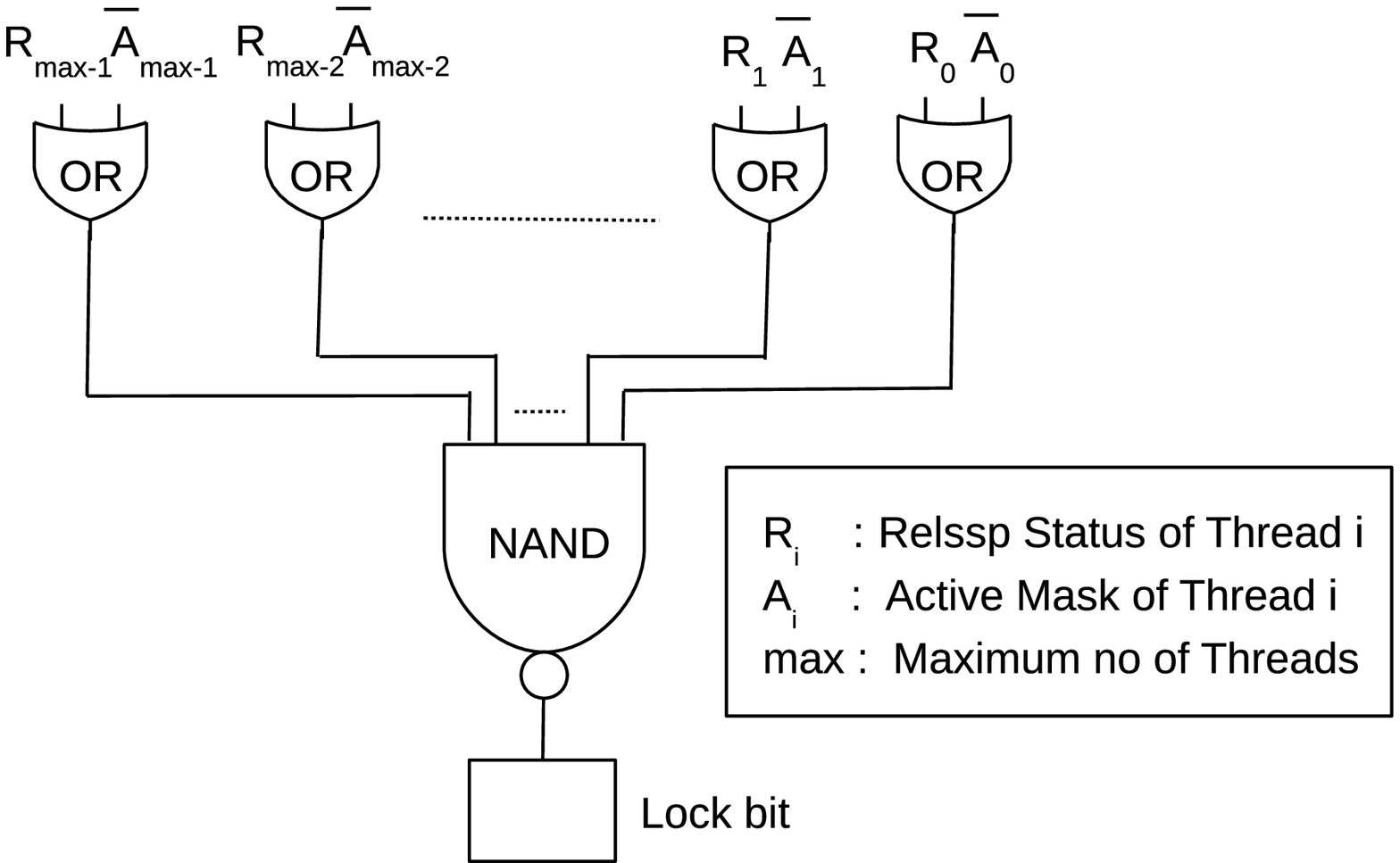}}
      {\caption{Hardware Implementation of \Release\ Instruction}\label{fig:circuit}}
\end{figure}

To minimize the delay in releasing the shared scratchpad, we propose a
new instruction, called \Release, in PTX assembly language. The
semantics of \Release\ instruction is to unlock the shared region only
when all active threads within a thread block finished executing the
shared region. Figure~\ref{clear} shows the pseudo code for
\Release\ instruction. The {\sc ReleaseSSP()} procedure maintains {\tt
  \relcnt}, an integer initialized to zero. When an active thread
within a thread block executes a \Release\ instruction, it increments
the {\tt \relcnt} value.
When all active threads of a thread block execute
\Release\ instruction (Line~\ref{done}, when {\tt
  \relcnt} equals {\tt ACTIVE\_THREADS}), the shared region is
unlocked by invoking {\sc UnlockSharedRegion()}. The unlock procedure
releases the shared scratchpad region by resetting the lock
variable. The execution of \Release\ by a thread block that does not
access shared scratchpad region has no effect.

It is clear that {\tt \relcnt} in Figure~\ref{clear} has to be a
shared variable, hence a software implementation will require to
manage critical section.  The same algorithm, however, can be
efficiently implemented in hardware circuit as shown in
Figure~\ref{fig:circuit}. The $i^{th}$ thread within a thread block is
associated with an active mask ($A_i$) and a release bit ($R_i$). The
mask $A_i$ is set if the $i^{th}$ thread is active. When this thread
executes \Release\ instruction, the release bit ($R_{i}$) gets
set. The shared scratchpad region is unlocked only when all the active
threads in a thread block execute \Release\ instruction (the lock bit,
i. e.\ the output of NAND gate becomes 0 in
Figure~\ref{fig:circuit}). In other words, shared scratchpad region is
unlocked if $\forall i$ $A_i \rightarrow R_i$ is true.

\subsection{Algorithm for optimal placement of \Release\ instruction} \label{subsec:optimal}

In Section~\ref{sec:release}, we introduced a new instruction to
release the shared scratchpad memory. In this section, we discuss a
compile-time analysis for optimal insertion of \Release\ instruction
in the program. We insert a \Release\ instruction at a program point
\pt\ such that the following conditions are met:

\begin{condition} \label{cond:1}
{\bf Safety:} The \Release\ instruction must be executed by
each active thread within a thread block, and it must be
executed after last access to shared scratchpad memory.
\end{condition}

\begin{condition}\label{cond:2}
{\bf Optimality:} The \Release\ instruction must be executed
by each active thread exactly once.
\end{condition}

Condition~\ref{cond:1}  ensures  that  shared  scratchpad  is
eventually released  by a thread block  since the instruction
is executed  by all the threads  of a thread block.  Also, it
guarantees that  shared scratchpad is released  only after a
thread    block   has    completed    using   it.    Whereas,
Condition~\ref{cond:2}   avoids    redundant   execution   of
\Release\ instruction.

In the scratchpad sharing, a thread block releases the shared
scratchpad memory after completing its execution, hence it is
equivalent to having a \Release\ instruction placed at the end of
the program, which guarantees both the conditions, albeit at the cost
of delay in releasing the shared scratchpad.  A simple improvement
that promotes early release of shared scratchpad memory and ensures
both the conditions, is to place the \Release\ instruction at a basic
block $BB_{postdom}$ where $BB_{postdom}$ is a common post dominator
of those basic blocks having the last accesses to the shared
scratchpad memory along different paths. Further, $BB_{postdom}$
should dominate \fgexit, i.e., it should be executed in all possible
execution paths. As the following example shows, this strategy, though
an improvement over placing \Release\ in \fgexit, may also result in
delaying the release of shared scratchpad memory.

\begin{figure}[t]
  \twofigure{\includegraphics[scale=0.35]{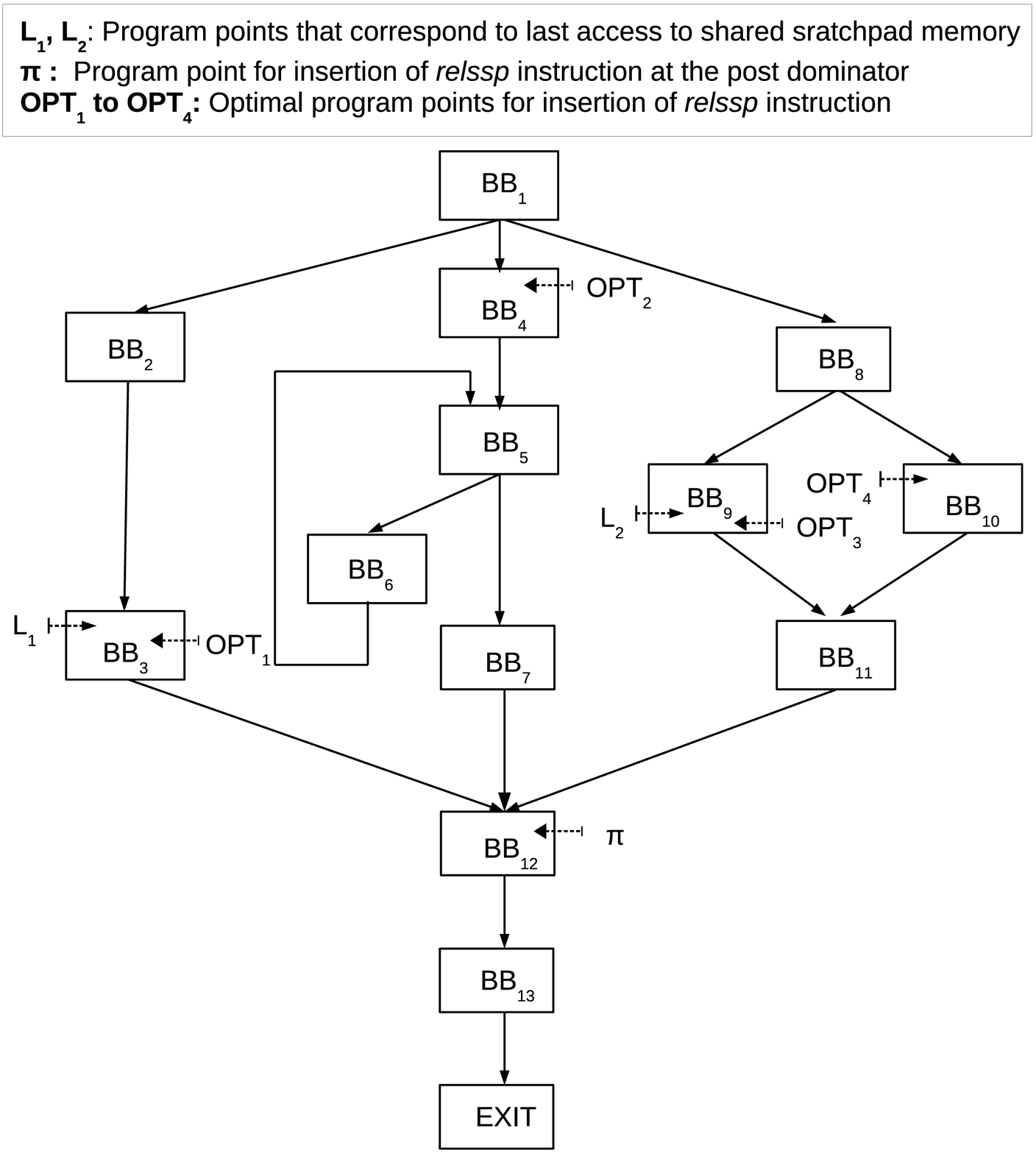}}
         {\vskip -2mm \caption{Possible Insertion Points for \Release} \label{fig:relssp}}
         {\includegraphics[scale=0.41]{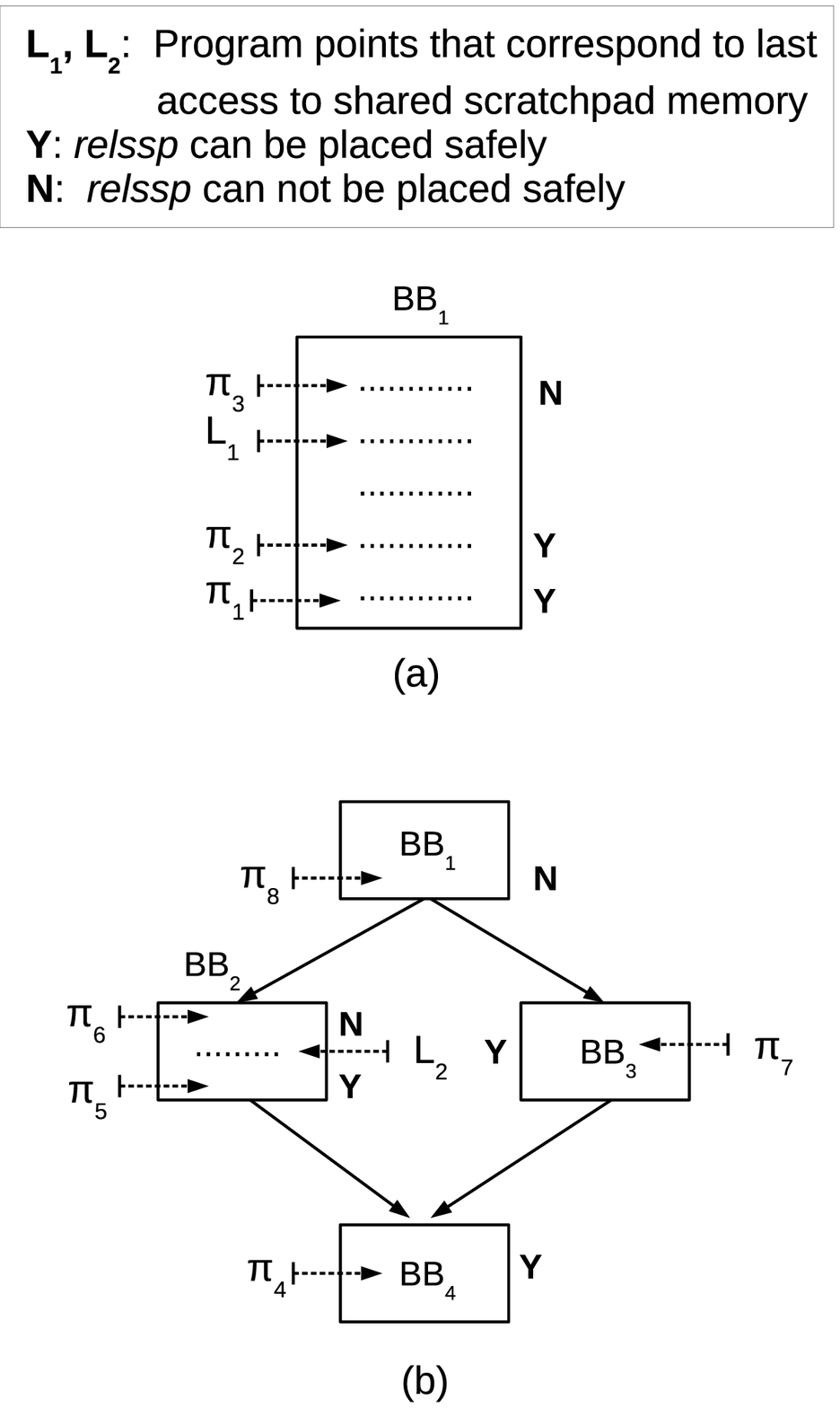}}
         {\vskip -2mm \caption{Scenarios for Optimal Insertion of \Release} \label{fig:meet}}
         \vskip -1mm
\end{figure}

\begin{example} \label{ex:postdom}
Consider a CFG shown in Figure~\ref{fig:relssp}.  Assume that $L_1,
L_2$ denote the program points that correspond to the last accesses to
shared scratchpad memory.  Since \Release\ instruction is to be
executed by all the threads of thread block, it can be placed at the
post dominator of the basic blocks $BB_3$ and $BB_9$, i.e., program
point marked \pt\ in $BB_{12}$, which is visible to all
threads. However, this delays the release of shared
scratchpad. Consider a thread that takes a path along $BB_9$, it can
execute \Release\ immediately after executing the last access to
shared region (shown as $OPT_3$ in the figure). It executes
\Release\ at program point \pt. Similarly, when a thread takes a path
along the basic block $BB_4$, it releases the shared scratchpad at
\pt\ even though it does not access any shared scratchpad in that
path. The scratchpad can be released at program point $OPT_2$
in $BB_4$.\hfill\qed
\end{example}

As is clear from the above example, placement of \Release\ instruction
has an effect on the availability of shared scratchpad memory.
Intuitively, a safely placed \Release\ instruction at a program point
\pt\ can be moved to a previous program point $\pt'$ in the same basic
block provided the intervening instructions do not access shared
scratchpad. The movement of \Release\ from a basic block $BB$ to
predecessor $BB'$ is possible provided every other successor of $BB'$
also does so.

\begin{example} \label{ex:opt1} 
Figure~\ref{fig:meet}(a) shows a basic block $BB_1$, which has the
last access to the shared scratchpad memory at $L_1$. In this block, if the
\Release\ instruction can be placed safely at the program point
$\pt_1$, then it can be moved to $\pt_2$ since there is no access to
shared scratchpad memory between $\pt_1$ and $\pt_2$. However, it can
not be moved to the program point $\pt_3$ within the same basic block,
because it violates safety (Condition~\ref{cond:1}).

Consider another scenario shown in Figure~\ref{fig:meet}(b), basic
block $BB_2$ has the last access to shared memory at $L_2$, and basic blocks
$BB_1$, $BB_3$, and $BB_4$ do not access any scratchpad memory. If the
\Release\ instruction can be placed safely at $\pt_4$ in $BB_4$, then
it can be moved to a program point $\pt_5$ and $\pt_7$ in the basic
blocks $BB_2$ and $BB_3$ respectively. However, it can not be moved to
program point $\pt_6$ in $BB_2$ and $\pt_8$ in $BB_1$ since it
violates of Condition~\ref{cond:1}. Also, the \Release\ instruction
can not be moved from $\pt_7$ in $BB_3$ to $\pt_8$ in $BB_1$ since the
basic block $BB_2$, which is a successor of $BB_1$, does not allow the
\Release\ instruction to be placed at $\pt_8$.\hfill\qed
\end{example}

\newcommand{\insafe}[1]{\ensuremath{{\sf SafeIN}(#1)}}
\newcommand{\outsafe}[1]{\ensuremath{{\sf SafeOUT}(#1)}}
\newcommand{\insertion}[1]{\ensuremath{{\sf INS}_{#1}}}

We now formalize these intuitions into a backward data flow analysis. The notations used are:

\begin{itemize}
\item \IN{BB} denotes the program point before the first statement of the basic block $BB$. \OUT{BB} denotes the program point after the last statement of $BB$.
\item \insafe{BB} is true if the \Release\ instruction can be safely placed at \IN{BB}, and  \outsafe{BB} is true if the \Release\ instruction can be safely placed at \OUT{BB}.
\item \insertion{\pi}, if true, denotes that \Release\ will be placed at program point $\pi$ by the analysis.
\end{itemize}

The data flow equations are:
\[\small\begin{array}{r@{\ }c@{\ }l@{\qquad}r@{\ }c@{\ }l}
  \insafe{BB} &=& \left\{\renewcommand{\arraystretch}{1.1}
  \begin{array}{l}
    \mbox{false, if $BB$ has shared}\\\qquad\;\;\mbox{scratchpad access}\\
    \outsafe{BB},\  \mbox{otherwise}
  \end{array}\right. &
  \outsafe{BB} &=& \left\{\renewcommand{\arraystretch}{1.1}
  \begin{array}{l}
    \mbox{true, if $BB$ is \fgexit\ block}\\
    \bigwedge\limits_{BS \in \Succ{BB}} \!\!\!\!\insafe{BS},\ \mbox{otherwise}
  \end{array}\right.
\end{array}\]

The above equations compute the  program points where \Release\ can be
placed safely.  For  a basic block $BB$, \OUT{BB} is  an optimal place
for  \Release\  instruction, if  \Release\  can  be placed  safely  at
\OUT{BB}, and it can not be moved safely to its previous program point
in the basic block, i.e., \IN{BB} is \isfalse.
This is computed as:
\begin{equation} \label{eq:out}
\small\insertion{\OUT{BB}} = \outsafe{BB}\wedge\neg(\insafe{BB})
\end{equation}

Similarly, \IN{BB} is an optimal point for \Release\ instruction, 
when the instruction can not be moved to its predecessors
\footnote{Absence of critical edges   guarantees that the instruction 
can either be moved to {\em     all} predecessors or to {\em none}.}. This
can be computed as:
\begin{equation} \label{eq:in}
\small\insertion{\IN{BB}} = \neg \left(\hspace*{-1mm}\bigwedge_{BP \in \Pred{BB}}  \hspace*{-5mm}\outsafe{BP}\hspace*{-1mm}\right) \wedge \insafe{BB}
\end{equation}

Equations~(\ref{eq:out})~and~(\ref{eq:in}) together, along with the absence of critical edges, ensure the optimality condition 
that each thread executes the \Release\ instruction exactly once.

\section{Requirements for Scratchpad Sharing} \label{sec:analysis}
\subsection{Hardware Requirements}

\begin{wrapfigure}{r}{0.5\textwidth}
  \includegraphics[scale=0.3]{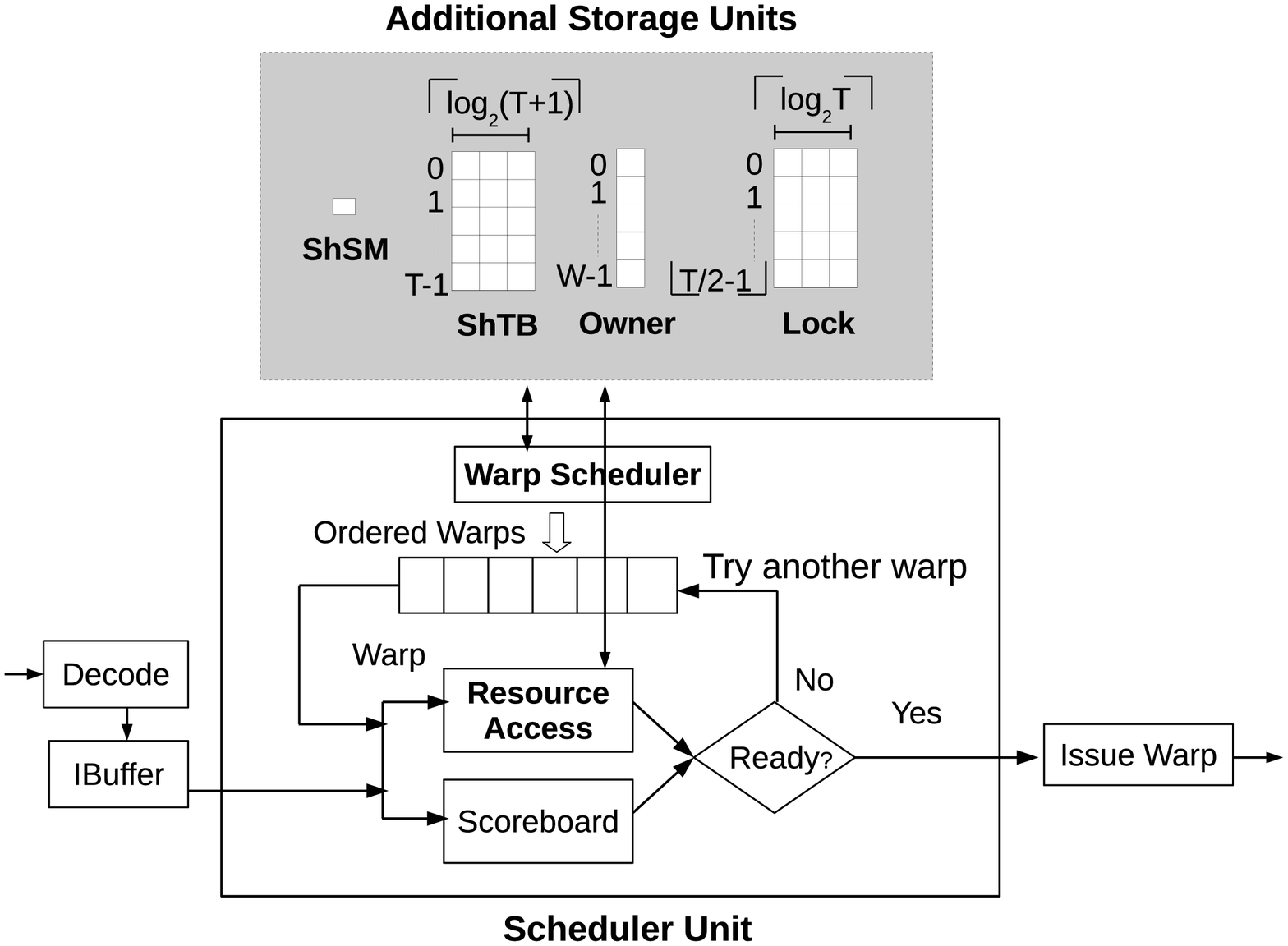}
  \caption{Modified Architecture for Scratchpad Sharing} \label{fig:modifiedGPU}
\end{wrapfigure}

Figure~\ref{fig:modifiedGPU} shows the modified GPU architecture to
implement scratchpad sharing.  Our approach requires two modifications
to scheduler unit in the SM. The first change is to the warp scheduler
which uses OWF optimization. The second change is the inclusion of
resource access unit, which follows the scratchpad access mechanism as
discussed in Section~\ref{sec:scratchpadsharing}. The resource
access unit requires the following additional storage:

\begin{enumerate}
\item Each SM requires a bit (shown as \emph{ShSM} in
  Figure~\ref{fig:modifiedGPU}) to indicate whether the scratchpad sharing is enabled for 
  it. This bit is set when the number of thread blocks
  launched using our approach is more than that of baseline approach.
\item Every thread block involved in sharing stores the id of its
  partner thread block in the \emph{ShTB} table. If a thread block is
  in unsharing mode, a $-1$ is stored. For $T$ thread blocks in
  the SM, we need a total of $T\log_2(T+1)$ bits.
\end{enumerate}

\begin{enumerate}
  \setcounter{enumi}{2}
\item Each warp a requires a bit to specify if it is an owner warp. For $W$ warps in the SM, 
$W$ bits are needed.  
  
\item For each pair of shared thread blocks, a lock variable is needed 
in order to access shared scratchpad memory.  This variable
  is set to the id the thread block that acquired shared scratchpad
  memory. For $T$ thread blocks, there are at most $\lfloor{T}/{2}\rfloor$ pairs of
  sharing thread blocks in the SM. This requires
  $\lfloor\frac{T}{2}\rfloor \lceil(\log_2T)\rceil$ bits in the SM.
\end{enumerate}

If a GPU has $N$ SMs and allows a maximum of $T$ thread blocks and $W$ warps per SM,
then the number of additional bits required is:
$\left(1+T\log_2(T+1)+W+\lfloor\frac{T}{2}\rfloor \lceil(\log_2T)\rceil\right)*N$. 
For the architecture we used for simulation (shown in Table~\ref{table:GPGPUArch}), 
the overhead is 209 bits per SM. In addition, each scheduler unit in the SM requires two comparator circuits and one arithmetic circuit to set the lock (See
Figure~\ref{fig:spaccess}).

\subsection{Analysis of Compiler Optimizations}
The dataflow analyses to compute definitions and usages of scratchpad
variables (Section~\ref{sec:minregion}) are bit-vector data flow
analyses~\cite{Khedker.DFA}. For a kernel with $n$ scratchpad variables and $m$
nodes (basic blocks) in the flow graph, the worst case complexity is
$O(n\times m^2)$ (assuming set operations on $n$ bit-wide vectors take
$O(n)$ time).

The computation of access ranges for sets of variables may require
analyzing all $O(2^n)$ subsets in the worst case, where the largest
size of a subset is $O(n)$. Thus, given the usage and definitions at
each program point in the kernel, computation of {\sf AccIN} and {\sf
  AccIN} requires $O(m\times n\times 2^n)$ time. Therefore, the total
time complexity is $O(n\times m^2 + m\times n\times 2^n)$.  Since the
number of scratchpad variables in a kernel function is small
(typically, $n \leq 10$), the overhead of the analysis is
practical.

Our approach inserts \Release\ instructions in a CFG such that
\Release\ is called exactly once along any execution path. In the
worst case, all nodes in a CFG (except \fgentry\ and \fgexit\ blocks)
might fall along different paths from \fgentry\ to \fgexit. Hence the
worst case number of \Release\ inserted is $O(m)$.

\section{Experimental Evaluation}\label{sec:experimentanalysis} 

\begin{table}[t]
\caption{Set-3 Benchmarks: The Number of Thread Blocks is Not Limited by Scratchpad Memory}\label{table:set_3}
\vskip -2mm 
\centering
\scalebox{0.85}{\renewcommand{\arraystretch}{1.05}
\begin{tabular}{l@{\quad }l@{\quad}l@{\quad}l@{\ \ }l@{\ \ }}
\hline\hline
Benchmark &  Application & Kernel &  Limited by\\
\hline
GPGPU-SIM & BFS & Kernel & Threads, Registers \\
RODINIA & b+tree & findRangeK & Registers \\
CUDA-SDK & dct8x8\_5 (DCT5) & CUDAkernel1DCT & Blocks \\
RODINIA & gaussian & FAN1 & Threads \\
GPGPU-SIM & NN & executeSecondLayer & Blocks \\
\hline
\end{tabular}}
\vskip -2mm
\end{table}

\begin{table}[t]
\centering
\caption{Details of Modifications to the Benchmarks} \vskip -3mm
\label{tab:modifications}
{\scalebox{0.85}{\renewcommand{\arraystretch}{1.0}
\begin{tabular}{llllll}
\hline \hline
Application & File Name & Line Number & \begin{tabular}[c]{@{}l@{}}Constant/Variable Name\\ \end{tabular} & Old Value & New Value \\ \hline
backprop & backprop.h & \begin{tabular}[c]{@{}l@{}}10\\ 11\end{tabular} & \begin{tabular}[c]{@{}l@{}}WIDTH\\ HEIGHT\end{tabular} & \begin{tabular}[c]{@{}l@{}}16\\ 16\end{tabular} & \begin{tabular}[c]{@{}l@{}}48\\ 48\end{tabular} \\ \hline
DCT[1..4] & dct8x8.cu & 317 & numIterations & 100 & 1 \\ \hline
NQU & nqueen.c & 11 & THREAD\_NUM & 96 & 64 \\ \hline
SRAD1 & \multirow{2}{*}{srad.h} & \multirow{2}{*}{3} & \multirow{2}{*}{BLOCK\_SIZE} & \multirow{2}{*}{16} & \multirow{2}{*}{24} \\ 
SRAD2 &  &  &  &  &  \\ \hline
heartwall & define.c & \begin{tabular}[c]{@{}l@{}}9\\ 13\end{tabular} & \begin{tabular}[c]{@{}l@{}}NUMBER\_THREADS\\ ALL\_POINTS\end{tabular} & \begin{tabular}[c]{@{}l@{}}512\\  51\end{tabular} & \begin{tabular}[c]{@{}l@{}}128\\ 140\end{tabular} \\ 
   & main.cu \tablefootnote{This has been changed for quick simulation.} & 96 & frame\_processed & 5 & 1 \\ \hline
histogram & histogram\_ & 18 & HISTOGRAM64\_ & 256 & 384 \\ 
		  & common.h 	 & 	  & BIN\_COUNT 	   & 	 & \\ \hline
NW1 & \multirow{2}{*}{needle.h} & \multirow{2}{*}{1} & \multirow{2}{*}{BLOCK\_SIZE} & \multirow{2}{*}{32} & \multirow{2}{*}{16} \\ 
NW2 &  &  &  &  &  \\ \hline
\end{tabular}}}
\vskip -2mm
\end{table}

We implemented the proposed scratchpad sharing approach and integrated
\Release\ instruction in GPGPU-Sim V3.x~\cite{GPGPUSIM} simulator. We
implemented the compiler optimizations in PTX assembly~\cite{PTX}
using Ocelot~\cite{Ocelot} framework. The baseline architecture that
we used for comparing our approach is shown in
Table~\ref{table:GPGPUArch}. We evaluated our approach on several
kernels from CUDA-SDK~\cite{CUDA-SDK}, GPGPU-Sim~\cite{GPGPU-Sim}, and
Rodinia~\cite{rodinia} benchmark suites.

Depending on the amount and the last usage of the shared scratchpad
memory by the applications, we divided the benchmark applications into
three sets.  Set-1 and Set-2 (Table~\ref{table:set_1}) consists of
applications whose number of resident thread blocks are limited by
scratchpad memory.  For Set-1, the applications do not access
scratchpad memory till towards the end of their execution, while for
Set-2, the applications access scratchpad memory till towards the end
of their execution.  The introduction of \Release\ instruction is
expected to give benefit over our earlier
approach~\cite{resourcesharing} only for Set-1 applications.  Set-3
benchmarks (Table~\ref{table:set_3}) consist of applications whose
number of thread blocks are not limited by scratchpad memory, but by
some other parameter. These are included to show that our approach
does not negatively affect the performance of applications that are
not limited by scratchpad memory.

For each application in Set-1 and Set-2 benchmarks,
Table~\ref{table:set_1} shows the kernel that is used for evaluation,
the number of the scratchpad variables declared in each kernel, the
amount of the scratchpad memory required for each thread block, and
the thread block size. Some applications in Set-1 and Set-2 benchmarks
are modified to make sure that the number of thread blocks is limited
by scratchpad memory, thus making scratchpad sharing approach
applicable. These changes increase the scratchpad memory requirement
per thread block and are shown in
Table~\ref{tab:modifications}.  For Set-3 benchmarks,
Table~\ref{table:set_3} shows the cause of limitation on the number of
thread blocks. The causes include the limit on the number of
registers, the maximum limit on the number of resident thread blocks,
and the maximum limit on the number of resident threads.

We compiled all the applications using CUDA 4.0\footnote{GPGPU-Sim and
  Ocelot do not support CUDA 5.0 and above.}  and simulated them using
the GPGPU-Sim simulator.  We use a threshold ($t$) to configure the
amount of scratchpad sharing. If each thread block requires $R_{tb}$
amount of scratchpad memory, then we allocate $R_{tb}(1+t)$ for each
pair of shared thread blocks, in which we allocate $R_{tb}(1-t)$ as
shared scratchpad memory. We analyzed the benchmark applications for
various threshold values and choose the value $t$ as 0.1 (i.e., 90\%
scratchpad is shared among pair of thread blocks) to give the maximum
benefit. The details of the experiments to choose $t$ are given in
technical report~\cite{resourcesharingarxiv}.

We measure the performance of our approach using the following metrics:
\begin{enumerate}
\item The {\bf number of the resident thread blocks} launched in the
  SMs. This is a measure of the amount of thread level parallelism
  present in the SMs.
\item The {\bf number of instructions executed per shader core clock
  cycle (IPC)}. This is a measure of the throughput of the GPU
  architecture.
\item The {\bf number of simulation cycles} that an application takes
  to complete its execution. This is a measure of the performance of
  the benchmark applications.
\end{enumerate}

\subsection{Analysis of Set-1 and Set-2 Benchmarks}
We use \emph{Unshared-LRR} to denote the baseline unsharing approach,
\emph{Shared-OWF} to denote our scratchpad sharing approach with OWF
scheduler, and \emph{Shared-OWF-OPT} to denote the scratchpad sharing
approach that includes OWF scheduler and compiler optimizations.

\begin{figure}
	\centering
  	\includegraphics[scale=0.5]{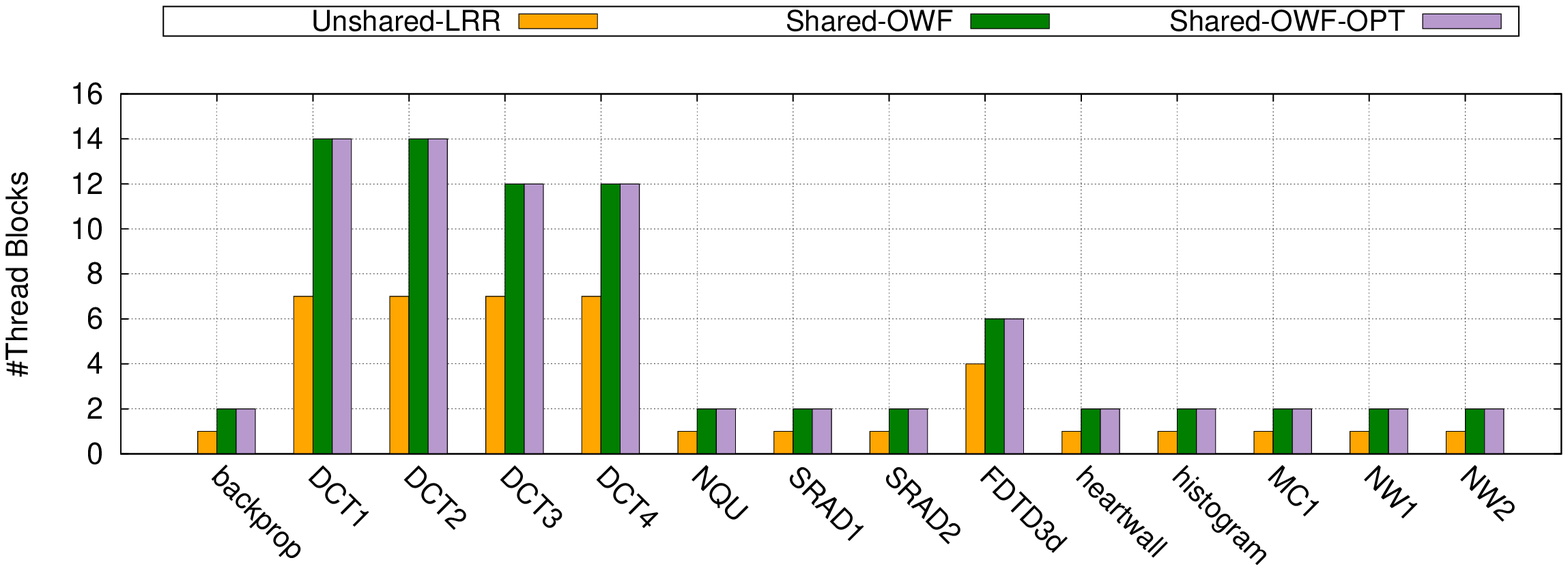}
	\vskip -3mm
	\caption{Comparing the Number of Resident Thread Blocks}\label{fig:blocks}
	\vskip -3mm
\end{figure}

\subsubsection{Comparing the Number of Resident Thread Blocks}
Figure~\ref{fig:blocks} shows the number of thread blocks for the three
approaches. For  applications \emph{DCT1} and
\emph{DCT2}, \emph{Unshared-LRR} launches 7 thread blocks in the SM
according to the amount of scratchpad memory required by their thread
blocks. \emph{Shared-OWF} launches 14 thread blocks in the SM, where
each of the 7 additional thread blocks share scratchpad memory with
other resident thread blocks.  For \emph{DCT3} and \emph{DCT4}
applications, \emph{Unshared-LRR} launches 7 thread blocks in the SM,
whereas \emph{Shared-OWF} launches 12 thread blocks in the SM such
that the additional 5 thread blocks share scratchpad memory with the
existing 5 thread blocks; while the remaining 2 existing thread blocks
in the SM do not share scratchpad memory with any other thread block.
For \emph{FDTD3d}, \emph{Shared-OWF} launches 2 additional thread
blocks in the SM when compared to \emph{Unshared-LRR}, which share
scratchpad memory with other 2 resident thread blocks. For the
remaining applications, \emph{Unshared-LRR} launches 1 thread block,
whereas \emph{Shared-OWF} launches 1 additional thread block in the SM
which shares scratchpad memory with the existing thread block.  Note
that the number of thread blocks launched by \emph{Shared-OWF-OPT} is
exactly same as that of \emph{Shared-OWF}. This is expected since the
number of additional thread blocks launched by scratchpad sharing
approach depends on two parameters: (1) the amount of scratchpad
sharing, and (2) the amount of scratchpad memory required by a thread
block; and our compiler optimizations do not affect either of these
parameters.

\subsubsection{Performance Comparison}
\begin{figure}[t]
{\includegraphics[scale=0.5]{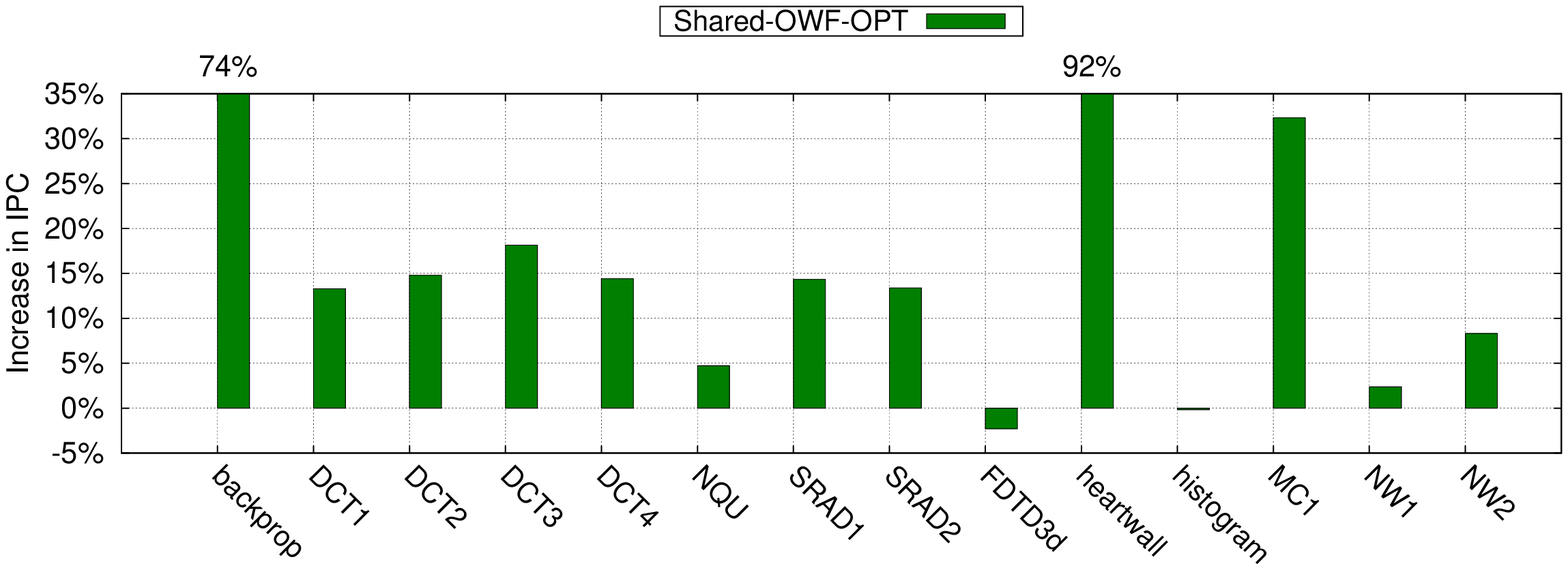}}
	\vskip -3mm
	\caption{Comparing the IPC}\label{fig:ipc}
	\vskip -3mm          
\end{figure}

Figure~\ref{fig:ipc} compares the performance of \emph{Shared-OWF-OPT}\footnote{IPC for \emph{Shared-OWF-OPT} also takes into account the extra instructions inserted by the compiler optimizations. The absolute IPC values are shown in Table~\ref{table:ipc} in Appendix~\ref{sec:appendix}.}
in terms of the number of instructions executed per cycle (IPC) with
that of \emph{Unshared-LRR}. 
We observe a maximum improvement of
92.17\% and an average (Geometric Mean) improvement of 19\% with
\emph{Shared-OWF-OPT}. The maximum benefit of 92.17\% is for
\emph{heartwall} because the additional thread blocks launched by
\emph{Shared-OWF-OPT} do not access the shared scratchpad
region. Hence all the additional thread blocks make progress without
waiting for corresponding shared thread blocks. \emph{MC1} improves by
32.32\% because additional thread blocks launched in the SM make
significant progress before accessing shared scratchpad
region. \emph{backprop} shows an improvement of 74.2\%, it leverages
both scratchpad sharing and the compiler optimizations to perform
better. The improvements in \emph{SRAD1} and \emph{SRAD2} applications
are largely due to the compiler optimizations. \emph{FDTD3d} slows
down (--2.29\%) with \emph{Shared-OWF-OPT} due to more number of L1
and L2 cache misses when compared to
\emph{Unshared-LRR}. \emph{histogram} does not benefit from sharing
since the thread blocks start accessing shared scratchpad region early
in the execution, causing one of the blocks from each sharing pair to
wait for the lock.

\begin{table}[t]
  \caption{Comparing the Number of Simulated Instructions}  
\vskip -2mm 
\centering
\scalebox{0.75}{\renewcommand{\arraystretch}{1.05}
\begin{tabular}{@{}l@{ }r@{\quad}r@{ \quad}r@{\quad}r@{\quad}r@{}}
\hline\hline
Benchmark & Threads & Unshared-LRR (U) & Shared-OWF (S)  & Shared-OWF-OPT (SO) & Difference (SO - U) \\
\hline
backprop & 1,048,576 & 131,203,072 & 131,203,072 & 133,234,688 & 2,031,616 \\
DCT1 & 32,768 & 9,371,648 & 9,371,648 & 9,404,416 & 32,768 \\
DCT2 & 32,768 & 9,502,720 & 9,502,720 & 9,535,488 & 32,768 \\
DCT3 & 32,768 & 11,255,808 & 11,255,808 & 11,304,960 & 49,152 \\
DCT4 & 32,768 & 11,157,504 & 11,157,504 & 11,206,656 & 49,152 \\
NQU & 24,576 & 1,282,747 & 1,282,747 & 1,331,515 & 48,768 \\
SRAD1 & 4,161,600 & 756,433,955 & 756,433,955 & 760,595,555 & 4,161,600 \\
SRAD2 & 4,161,600 & 450,077,975 & 450,077,975 & 454,239,575 & 4,161,600 \\
FDTD3d & 144,384 & 5,549,531,392 & 5,549,531,392 & 5,549,820,160 & 288,768 \\
heartwall & 17,920 & 11,280,920 & 11,280,920 & 11,316,760 & 35,840 \\
histogram & 46,080 & 893,769,168 & 893,769,168 & 893,861,328 & 92,160 \\
MC1 & 3,008 & 2,881,568 & 2,881,568 & 2,887,584 & 6,016 \\
NW1 & 3,184  & 5,580,458 & 5,580,458 & 5,583,642 & 3,184 \\
NW2 & 3,168 & 5,561,919 & 5,561,919 & 5,565,087 & 3,168 \\
\hline
\end{tabular}}
\label{table:insn}
\vskip -2mm
\end{table}

\subsubsection{Overhead of \Release\ Instruction}
Table~\ref{table:insn} shows the run-time overhead of inserting
\Release\ instruction.  We report sum of the number of instructions
executed by all threads for \emph{Unshared-LRR}, \emph{Shared-OWF},
and \emph{Shared-OWF-OPT}. We also report the number of threads
launched.

From the table, we observe that the number of instructions executed by
\emph{Unshared-LRR} and \emph{Shared-OWF} is same. This is because
\emph{Shared-OWF} does not insert \Release\/ instruction, and
hence the input PTX assembly is not altered. \emph{Shared-OWF-OPT}
increases number of executed instructions as it inserts \Release\/
and, in some cases, \emph{GOTO} instruction to split critical
edges. For the applications \emph{DCT1}, \emph{DCT2}, \emph{SRAD1},
\emph{SRAD2}, \emph{NW1}, and \emph{NW2}, the number of additionally
executed instructions (shown as \emph{Difference} (SO-U) in the table)
is equal to number of threads because \emph{Shared-OWF-OPT}
inserts only the \Release\ instruction. Further, each thread
executes \Release\ exactly once. For \emph{FDTD3d}, \emph{heartwall}, \emph{histogram}, and \emph{MC1} applications, the number of additional
instructions executed by \emph{Shared-OWF-OPT} is twice that of number
of threads. For these applications, each thread executes two
additional instructions, i.e., one \Release\ instruction, and one
\emph{GOTO} instruction for splitting a critical
edge.  For \emph{backprop}, \emph{DCT3}, \emph{DCT4}, and \emph{NQU} applications, some threads 
take a path that has two additional instructions ($GOTO$ and \Release), 
while other threads take the path which has one additional 
\Release\ instruction.

\begin{figure}
\centering
  	\includegraphics[scale=0.5]{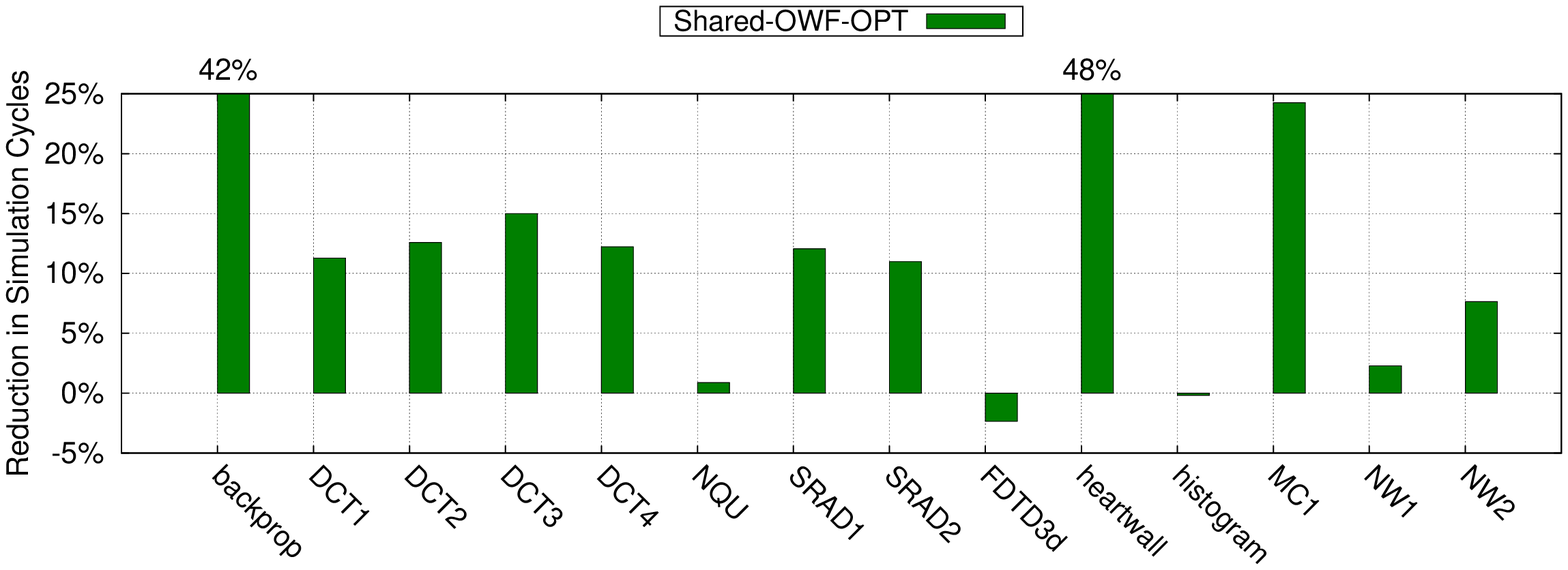}\vskip 1mm
             \caption{Comparing the Number of Simulation Cycles}\label{fig:simcycles}
\vskip -7mm
\end{figure}

\subsubsection{Reduction in Simulation Cycles}
Figure~\ref{fig:simcycles} shows the effectiveness of
\emph{Shared-OWF-OPT} by comparing the number of simulation cycles
with that of \emph{Unshared-LRR}. We observe a maximum reduction of
47.8\% and an average reduction of 15.42\% in the number of
simulation cycles when compared to \emph{Unshared-LRR}. Recall that
\emph{Shared-OWF-OPT} causes applications to execute more number of
instructions (Table~\ref{table:insn}). These extra instructions are also
counted while computing the simulation cycles for \emph{Shared-OWF-OPT}.

\subsubsection{Effectiveness of Optimizations}
\begin{figure}
\centering
\includegraphics[scale=0.53]{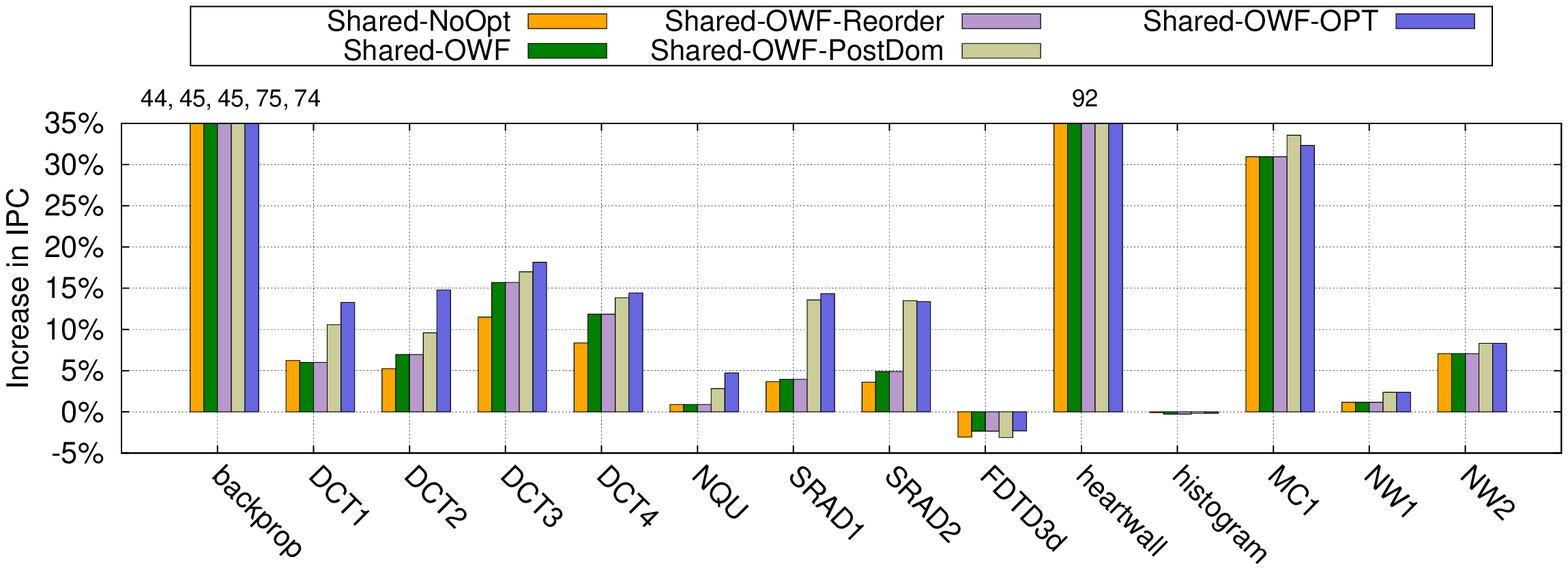}
\vskip 1mm
\caption{Performance Analysis of Optimizations}
\label{fig:optimizations}
\vskip -7mm
\end{figure}

Figure~\ref{fig:optimizations} shows the effectiveness of our
optimizations with scratchpad sharing. We observe that all
applications, except \emph{FDTD-3d} and \emph{histogram}, show some
benefit with scratchpad sharing even without any optimizations (shown
as \emph{Shared-NoOpt} in the figure). With OWF scheduling
(\emph{Shared-OWF}), applications improve further because OWF
schedules the resident warps in a way that the non-owner warps help in
hiding long execution latencies.  For our benchmarks, minimizing
shared scratchpad region (shown as \emph{Shared-OWF-Reorder}) does not
have any noticeable impact. This is because (a) Most applications
declare only a single scratchpad variable (Table~\ref{table:set_1}) in
their kernel, hence the optimization is not applicable (there is only
one possible order of scratchpad variable declarations); and (b) For
the remaining applications, the scratchpad declarations are already
ordered in the optimal fashion, i.e., the access to shared scratchpad
region is already minimal.

The addition of \Release\ instruction at the postdominator and at the
optimal places is denoted as \emph{Shared-OWF-PostDom} and
\emph{Shared-OWF-OPT} respectively. All Set-1 applications improve
with either of these optimizations because the \Release\ instruction
helps in releasing the shared scratchpad memory earlier. For
\emph{backprop} and \emph{SRAD2} applications,
\emph{Shared-OWF-PostDom} is better than \emph{Shared-OWF-OPT} because
the threads in \emph{backprop} execute one additional \emph{GOTO}
instruction with \emph{Shared-OWF-OPT} (\emph{Shared-OWF-PostDom} does
not require critical edge splitting). \emph{SRAD2} has more number of
stall cycles with \emph{Shared-OWF-Opt} as compared to
\emph{Shared-OWF-PostDom}. For most of the other benchmarks,
\emph{Shared-OWF-Opt} performs better as it can push \Release\/
instruction earlier than with \emph{Shared-OWF-PostDom}, thus
releasing shared scratchpad earlier allowing for more thread level parallelism.

As expected, Set-2 applications do not show much benefit with
\emph{Shared-OWF-PostDom} or \emph{Shared-OWF-OPT} since they access
shared scratchpad memory till towards the end of their
execution. Hence both the optimizations insert \Release\ instruction
in the \fgexit\ block in the CFGs. The application \emph{heartwall}
does not use shared scratchpad memory and hence it shows maximum
benefit even without the insertion of \Release\ instruction.

\subsubsection{Progress of Shared Thread Blocks}
\begin{figure}
\centering
\includegraphics[scale=0.53]{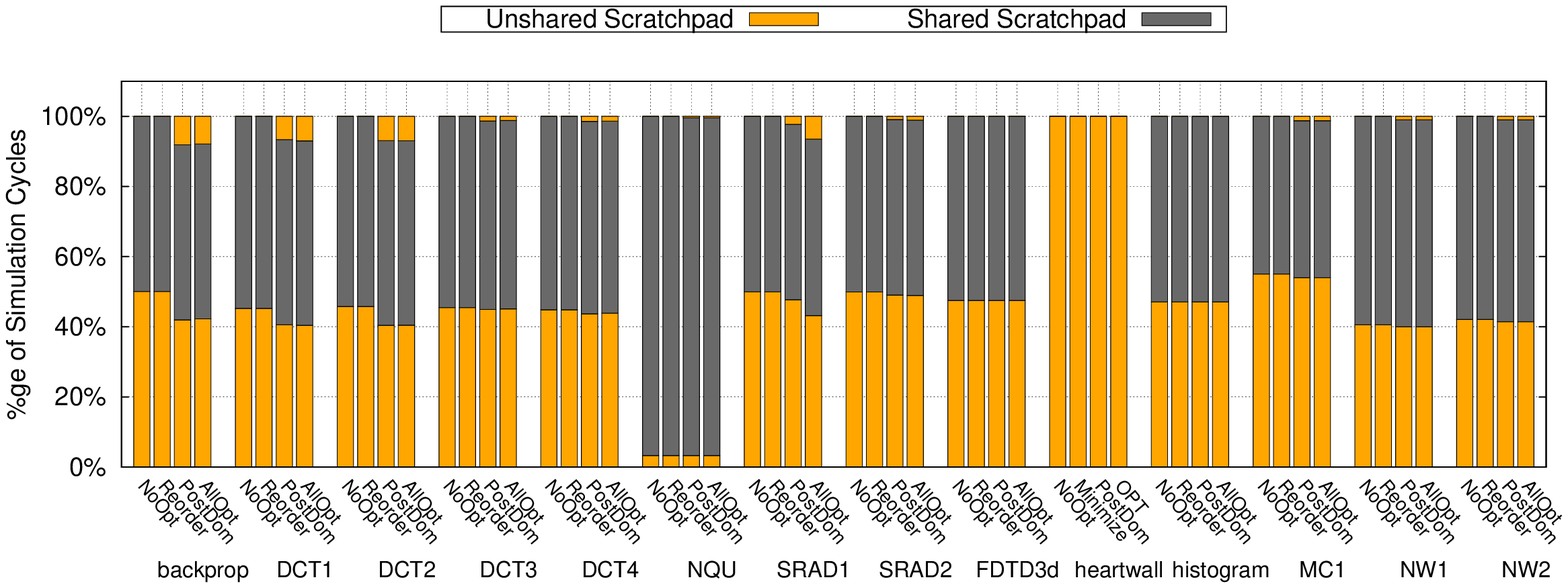}
\caption{Progress of Shared Thread Blocks}
\label{fig:progress}
\vskip -3mm
\end{figure}

Figure~\ref{fig:progress} shows the effect of compiler optimizations
by analyzing the progress of shared thread blocks through shared and
unshared scratchpad regions. In the figure, \emph{NoOpt} denotes the
default scratchpad sharing mechanism where none of our optimizations
are applied on an input kernel. \emph{Minimize} denotes the scratchpad
sharing approach which executes an input kernel having minimum access
to shared scratchpad region. \emph{PostDom} and \emph{OPT} use our
modified scratchpad sharing approach that execute an input kernel with
additional \Release\ instructions placed at post dominator and optimal
places (Section~\ref{subsec:optimal}) respectively. In the figure, we
show the percentage of simulation cycles spent in unshared scratchpad
region (before acquiring shared scratchpad), shared scratchpad region,
and unshared scratchpad region again (after releasing the shared
scratchpad) respectively.

\begin{figure}
\centering
\includegraphics[scale=0.5]{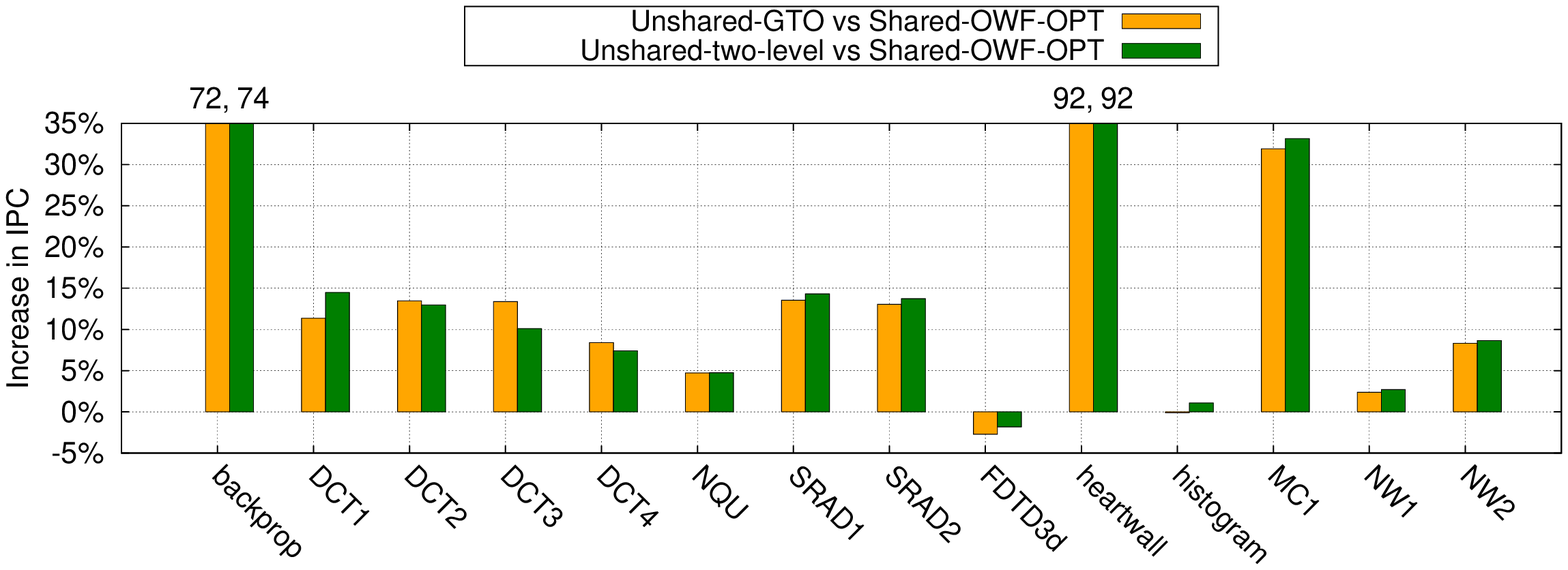}
\vskip -2mm
\caption{Improvement in IPC for \emph{Shared-OWF-OPT} w.r.t. baseline having (a) GTO scheduler, (b) two-level scheduler}\label{fig:2levelGTO}
\vskip -3mm
\end{figure}

From Figure~\ref{fig:progress} we observe that shared thread blocks in
all the applications access unshared scratchpad region before they
start accessing shared scratchpad memory. Hence all the shared thread
blocks can make some progress without wait. This progress is the main
reason for the improvements seen with scratchpad sharing
approach. Consider the application \emph{heartwall}, where none of the
shared thread blocks accesses shared scratchpad memory. Thus, all the
shared thread blocks in the application spend their execution in the
unshared scratchpad region. The compiler optimizations can not improve
the progress of shared thread blocks any further. \emph{Minimize} does
not affect \emph{DCT1}, \emph{DCT2}, \emph{DCT3}, \emph{DCT4},
\emph{FDTD3d}, \emph{histogram} applications because the kernels in
these applications declare single scratchpad variable. For the
remaining applications, \emph{Minimize} has same effect as that of
\emph{NoOpt}, because the default input PTX kernel already accesses
the shared scratchpad variables such that access to shared scratchpad
is minimum. We also observe that \emph{PostDom} and \emph{OPT}
approaches improve only those applications that spend considerable
simulation cycles in unshared scratchpad region \emph{after} last
access to shared scratchpad region.

\begin{figure}
\centering
\includegraphics[scale=0.5]{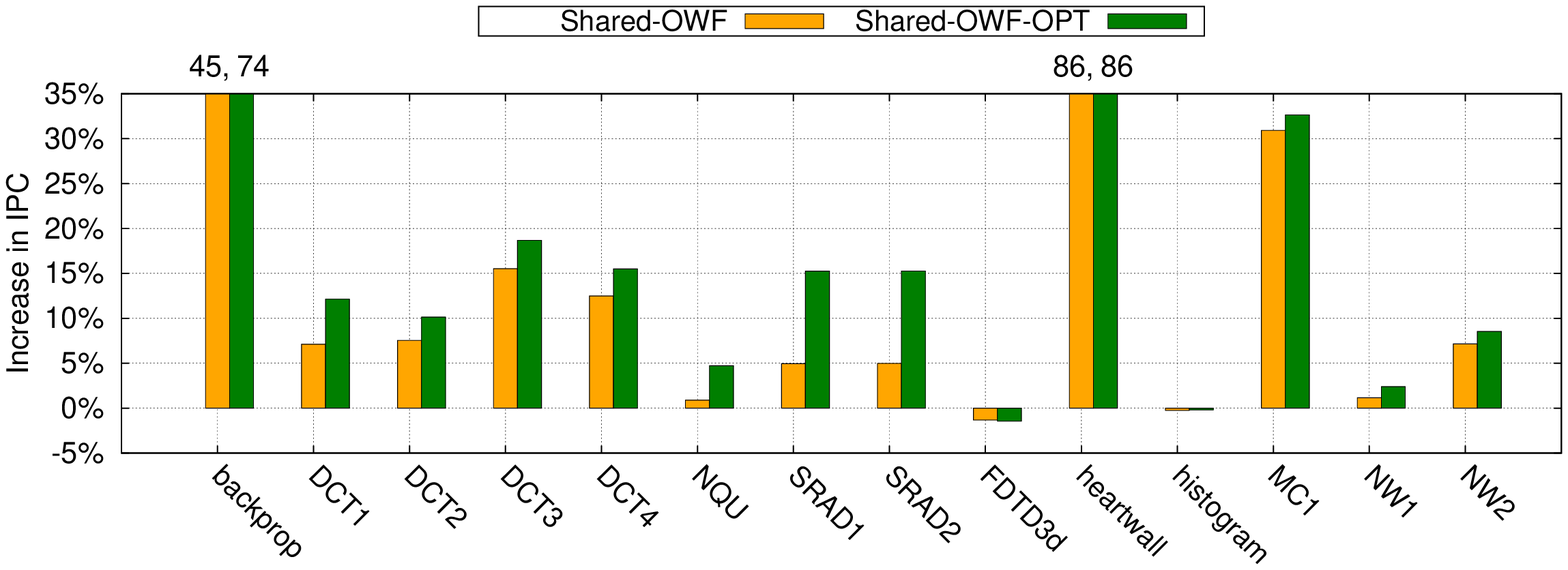}
\vskip -2mm
\caption{Performance Analysis with GPU Configuration: Scratchpad=16K, L1-Cache=48K}\label{fig:config1}
\vskip -3mm
\end{figure}

\begin{figure}
\centering
\includegraphics[scale=0.5]{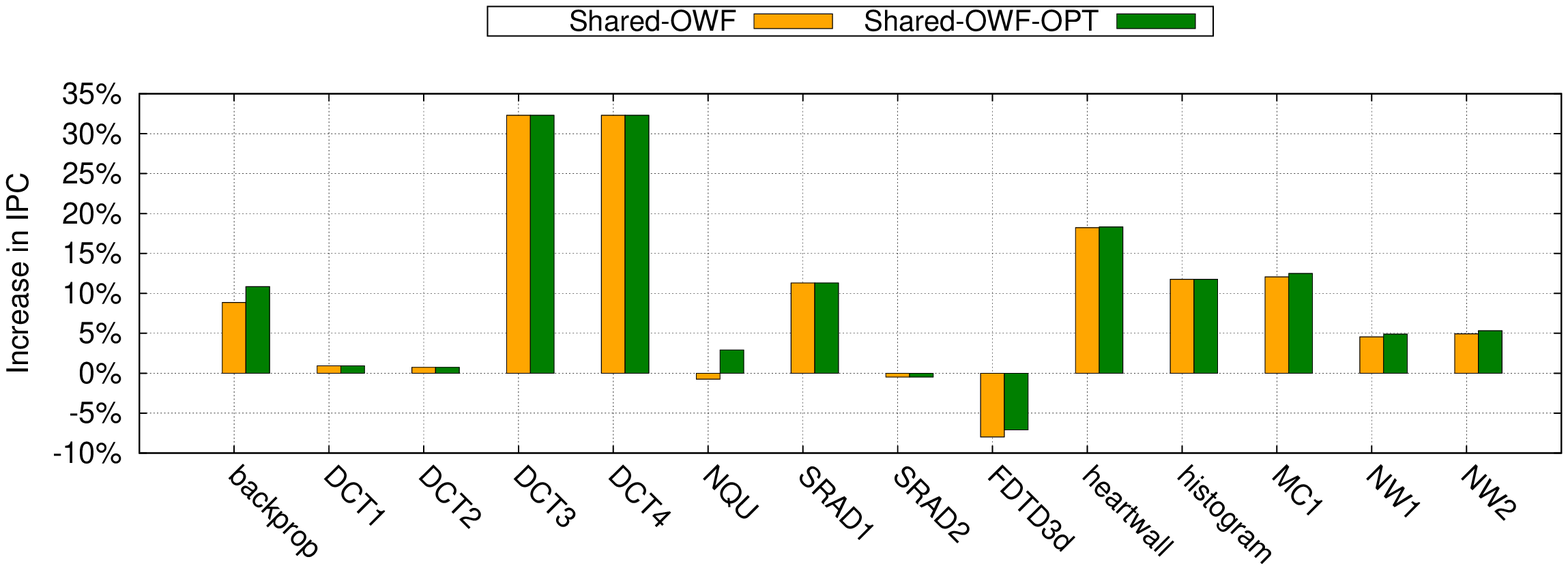}
\vskip -2mm
\caption{Performance Analysis with GPU Configuration: Scratchpad=48K, L1-Cache=16K, Resident Threads=2048}\label{fig:config2}
\vskip -3mm
\end{figure}

\subsubsection{Comparison with Different Schedulers}
Figure \ref{fig:2levelGTO} shows the effect of using
different scheduling policies.  The performance of
\emph{Shared-OWF-OPT} approach is compared with the baseline unshared
implementation that uses greedy then old (GTO) and two-level
scheduling policies respectively.  We observe that
\emph{Shared-OWF-OPT} approach shows an average improvement of 17.73\%
and 18.08\% with respect to unshared GTO and two-level scheduling
policies respectively. The application \emph{FDTD3d} degrade with our
approach when compared to the baseline with either GTO scheduling or
two-level scheduling since it has more number of L1 and L2 cache
misses with sharing. The application \emph{histogram} degrades with
sharing when compared to the baseline with GTO scheduling because of
more number of L1 misses. However \emph{histogram} with sharing
performs better than the baseline with two-level policy.

\subsubsection{Comparison with Other GPGPU-Sim Configurations}
We now compare the effectiveness of sharing approaches with
\emph{Unshared-LRR} for different GPGPU-Sim configurations.
Figure~\ref{fig:config1} shows results for a GPU configuration that uses
48K L1 cache.  We observe that sharing approach shows an average
improvement of 14.04\% with \emph{Shared-OWF} and 18.71\% with
\emph{Shared-OWF-OPT} over \emph{Unshared-LRR}. The
applications \emph{DCT3}, \emph{DCT4}, \emph{SRAD1}, \emph{SRAD2}, and
\emph{FDTD3d} using sharing are improved further with this
configuration since they benefit from increased L1 cache size. For \emph{heartwall}, \emph{Unshared-LRR}  benefits more with the increased L1 cache than the sharing approaches, it shows relatively less improvement of 86\% when compared to Figure~\ref{fig:ipc}.

Figure~\ref{fig:config2} shows the performance comparison of sharing
approaches with \emph{Unshared-LRR} for a GPU configuration that has
48K scratchpad memory and the maximum number of resident threads in
the SM as 2048.  We observe average improvements of 8.62\% and 9.21\%
with \emph{Shared-OWF} and \emph{Shared-OWF-OPT} approaches
respectively. Consider the applications \emph{DCT1}, \emph{DCT2},
\emph{DCT3}, and \emph{DCT4}. With increase in the scratchpad memory,
the number of resident thread blocks in the SM for these applications
is not limited by the scratchpad memory, hence sharing does not
increase the number of resident thread blocks. Also, the compiler
optimizations do not insert the \Release\ instruction into their PTX
code since there is no access to shared scratchpad region. Hence
\emph{Shared-OWF-OPT} behaves exactly same as \emph{Shared-OPT}. The
improvement in the performance of sharing approaches over
\emph{Unshared-LRR} is due to the OWF scheduling policy. OWF scheduler
arranges the resident warps according to the owner warps. Since all
the warps that are launched using \emph{Shared-OWF} own their
resources (no sharing), they become owner warps. Hence the warps are
arranged according to their dynamic warp id, giving the observed
benefit. While scratchpad sharing can increase the number of resident
thread blocks for \emph{SRAD1} and \emph{SRAD2}, no additional blocks
could be launched since the number of resident threads is restricted
to 2048.

\begin{figure}
\centering
\includegraphics[scale=0.5]{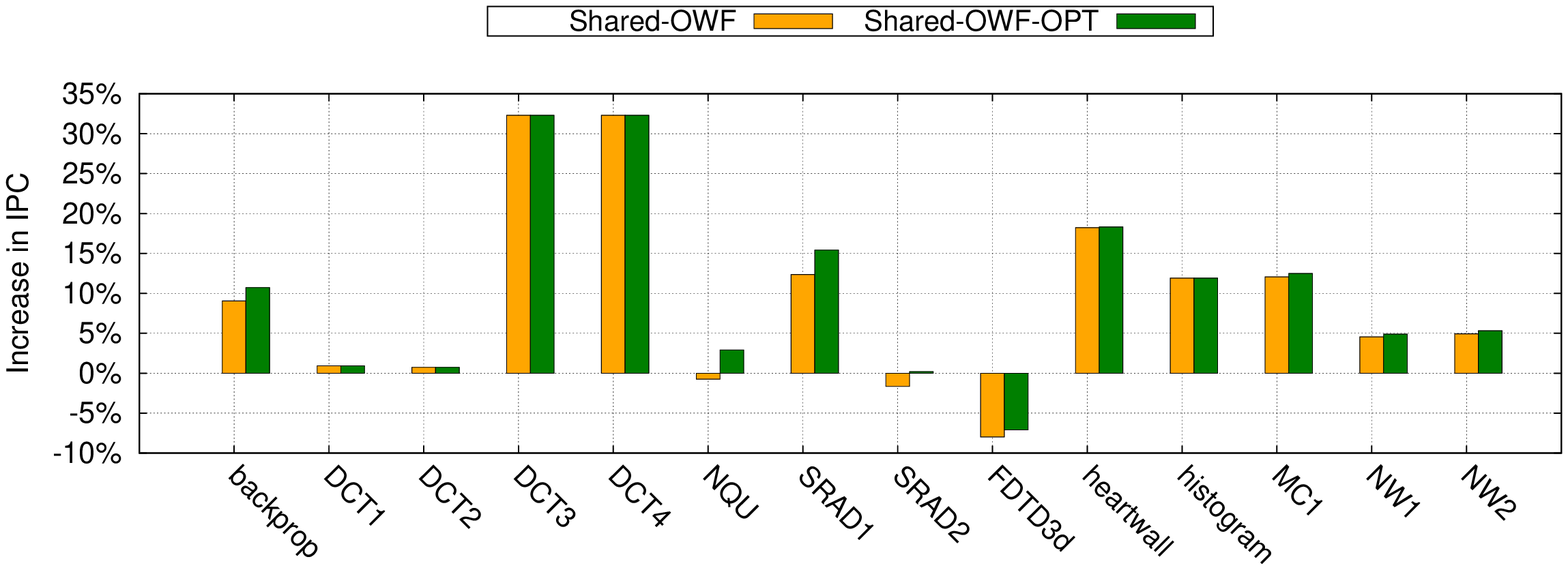}
\vskip -2mm
\caption{Performance Analysis with GPU Configuration: Scratchpad=48K, L1-Cache=16K, Resident Threads=3072}\label{fig:config3}
\vskip -3mm
\end{figure}

\begin{figure}
\centering
\includegraphics[scale=0.5]{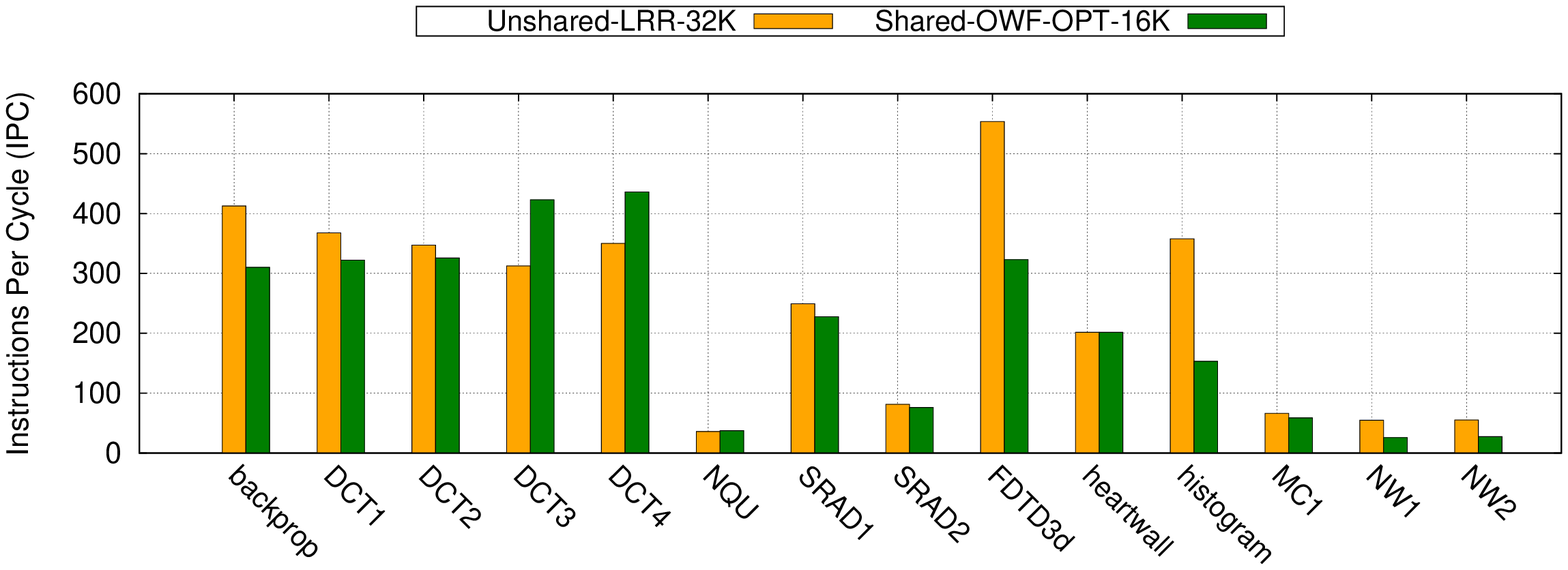}
\vskip -2mm
\caption{Comparison with Unshared-LRR that Uses Twice the Scratchpad Memory}\label{fig:resourcesavings}
\vskip -3mm
\end{figure}

Figure~\ref{fig:config3} shows the performance comparison for a GPU
configuration that uses 48K shared memory and the maximum number of
resident threads as 3072. With the number of resident threads
increasing from 2048 to 3072, sharing is able to increase the number
of resident thread blocks in \emph{SRAD1} and \emph{SRAD2}
applications, thereby improving the performance.

\subsubsection{Resource Savings}
Figure~\ref{fig:resourcesavings} compares the IPC of \emph{Shared-OWF-OPT} with \emph{Unshared-LRR} that uses twice the amount of
scratchpad memory on GPU. We
observe that \emph{DCT3}, \emph{DCT4}, \emph{NQU}, and
\emph{heartwall} show improvement with \emph{Shared-OWF-OPT} over \emph{Unshared-LRR} even with half the scratchpad memory. This is
because sharing helps in increasing the TLP by launching additional
thread blocks in each SM. The applications \emph{DCT1}, \emph{DCT2},
\emph{SRAD1}, \emph{SRAD2}, and \emph{MC1} applications perform
comparable with both the approaches. For the remaining applications,
\emph{Unshared-LRR} with double scratchpad memory performs better than
sharing since more number of thread blocks are able to make progress
with the former.

\begin{figure}
\centering
\includegraphics[scale=0.32]{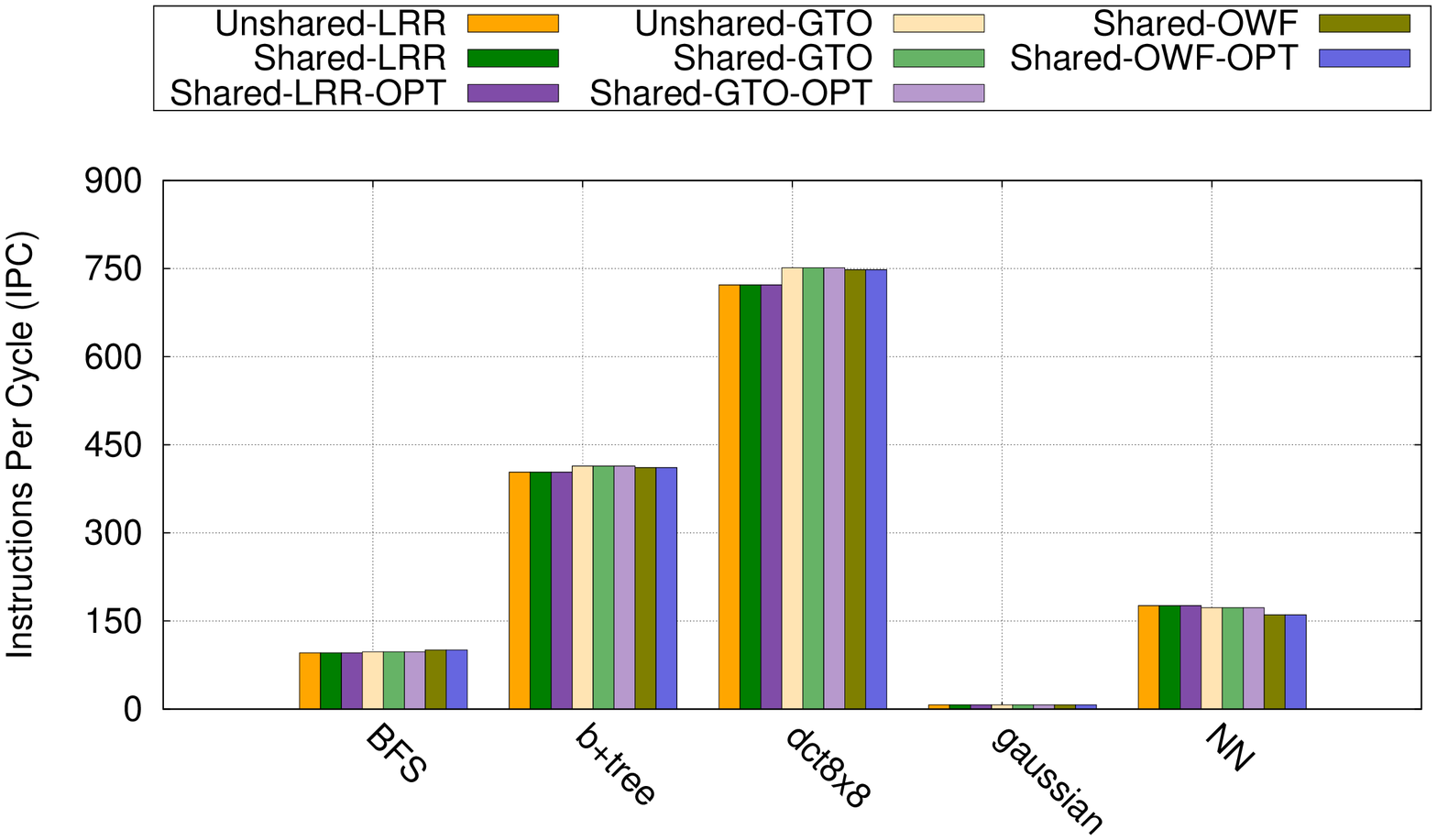}
\vskip -1mm
 \caption{Performance Comparison of Set-3 Benchmarks}\label{fig:set3}
\vskip -3mm
\end{figure}

\subsection{Analysis of Set-3 Benchmarks}
Performance analysis of Set-3 benchmarks is shown in the
Figure~\ref{fig:set3}. Recall that the number of thread blocks
launched by these applications is not limited by the scratchpad
memory. We observe that the performance of the applications with
\emph{Unshared-LRR}, \emph{Shared-LRR}, and \emph{Shared-LRR-OPT} is
exactly the same. For Set-3 applications all thread blocks are
launched in unsharing mode. Hence \emph{Shared-LRR} behaves exactly
same as \emph{Unshared-LRR}. Since these applications do not use any
shared scratchpad memory, our compiler optimizations do not insert
\Release\ instruction in their PTX code. Hence the number of
instructions executed by the \emph{Shared-LRR-OPT} approach is same as
that of \emph{Shared-LRR}. Similarly, we see that the performance of
applications with \emph{Unshared-GTO}, \emph{Shared-GTO}, and
\emph{Shared-GTO-OPT} is exactly the same. However, with OWF
optimization, \emph{Shared-OWF} and \emph{Shared-OWF-OPT} is
comparable to the \emph{Unshared-GTO} because OWF optimization
arranges the resident warps according to the owner. Since all the the
thread blocks own their scratchpad memory, they are sorted according
to the dynamic warp id. Hence they perform comparable to
\emph{Unshared-GTO}. The performances with \emph{Shared-OWF} and
\emph{Shared-OWF-OPT} are the same because the compiler optimizations
do not insert any \Release\ instruction.

\subsection{Additional Experiments}\label{sec:appendix}

\subsubsection{Performance Comparison with Other Configurations}
\begin{table}[t]
\centering
\setcounter{magicrownumbers}{0}
\newcommand\rownumber{\stepcounter{magicrownumbers}\arabic{magicrownumbers}}
\caption{Additional Benchmarks that are Limited by Scratchpad Memory} \label{table:set_4}
\vskip -2mm 
\centering
\scalebox{0.9}{\renewcommand{\arraystretch}{1.0}
\begin{tabular}{r@{\quad}l@{\ }l@{\quad}l@{\quad}r@{\ }r@{\ }r}
  \hline\hline
  & Benchmark &  Application & Kernel  & \#Scratchpad & Scratchpad   & Block \\
  &           &              &         & Variables    &  Size (Bytes)& Size  \\
\hline
\multicolumn{7}{c}{{\bf Benchmarks for 48KB / 64KB Scratchpad Memory}} \\ \hline
\rownumber. &  RODINIA & backprop & bpnn\_layerforward\_CUDA &2& 9408 & 256\\ 
\rownumber. &  CUDA-SDK & DCT1 & CUDAkernel2DCT  &1& 8320 & 128\\
\rownumber. &  CUDA-SDK & DCT2 & CUDAkernel2IDCT &1& 8320 & 128 \\
\rownumber. &  GPGPU-SIM & NQU & solve\_nqueen\_cuda\_kernel & 5 & 10496 & 64 \\
\rownumber. &  CUDA-SDK & histogram & histogram256Kernel & 1 & 9216 & 192 \\
\rownumber. &  CUDA-SDK & marchingCubes (MC2) & generateTriangles & 2 & 13824 & 48 \\
\rownumber. &  RODINIA & NW1 & needle\_cuda\_shared\_1 & 2 & 8452 & 32 \\
\rownumber. &  RODINIA & NW2 & needle\_cuda\_shared\_2 & 2 & 8452 & 32 \\
\hline
\multicolumn{7}{c}{{\bf Benchmarks for 48KB Scratchpad Memory}} \\ \hline
\rownumber. &  CUDA-SDK & FDTD3d & FiniteDifferencesKernel & 1 & 3840 & 128 \\
\rownumber. &  RODINIA & heartwall & kernel & 8 & 11872 & 128 \\
\rownumber. &  CUDA-SDK & marchingCubes (MC1) & generateTriangles & 2 & 9216 & 32\\\hline
\multicolumn{7}{c}{{\bf Additional Benchmarks for 16KB Scratchpad Memory}} \\ \hline
\rownumber. &  RODINIA & kmeans & kmeansPoint  & 2& 4608 & 576 \\
\rownumber. &  RODINIA & lud & lud\_internal & 2 & 3872 & 484 \\
\hline
\end{tabular}}
\end{table}

  To further verify the effectiveness of our approach, we
  evaluated it on two GPU configurations that use scratchpad
  memory of size 48KB and 64KB per SM
  (Table~\ref{table:KeplerMaxwell}). These configuration parameters are
  similar to that of NVIDIA's Kepler and Maxwell architectures
  respectively. The benchmarks that are used for the evaluation
  are shown in Table~\ref{table:set_4}. The changes in
  Table~\ref{table:set_4} with respect to those in
  Table~\ref{table:set_1} are:
  \begin{itemize}
  \item Kernel scratchpad memory size for \emph{DCT1} and
    \emph{DCT2} is increased from 2112 to 8320. This change
    ensures that applications are limited by scratchpad memory 
    for both the configurations. 
  \item A new application, \emph{MC2}, is created based on
    \emph{MC1}---the only difference being that kernel scratchpad
    memory size is increased to 13824, this is to enable scratchpad
    sharing for Configuration-2.
  \item The applications \emph{DCT3}, \emph{DCT4}, \emph{SRAD1},
    and \emph{SRAD2} are dropped as scratchpad sharing could not
    be made applicable even by increasing the kernel scratchpad
    memory size. These applications are limited either by the
    number of thread blocks or by the number of threads.
  \end{itemize}
  Note that scratchpad sharing is applicable for \emph{MC1},
  \emph{FDTD3d}, and \emph{heartwall} only with Configuration-1,
  but not with Configuration-2 (Table~\ref{table:KeplerMaxwell}).
  In addition, we performed experiments with 16KB scratchpad
  memory per SM for two new applications, \emph{kmeans} and
  \emph{lud}.

\begin{table}[t]
  \caption{GPGPU-Sim Configurations used for Additional Experiments}\label{table:KeplerMaxwell}
  \vskip -2mm 
  \scalebox{0.9}{\renewcommand{\arraystretch}{1.0}\begin{tabular}{l|r|r}
    \hline\hline
    Resource/Core & Configuration-1 & Configuration-2 \\          
    \hline
    Scratchpad Memory & 48KB & 64KB\\ 
    Max Number of Thread Blocks & 16 & 32 \\
    Max Number of Threads & 2048 & 2048\\ 
    \hline
  \end{tabular}}
\end{table}

\begin{figure}
\centering
\includegraphics[scale=0.4]{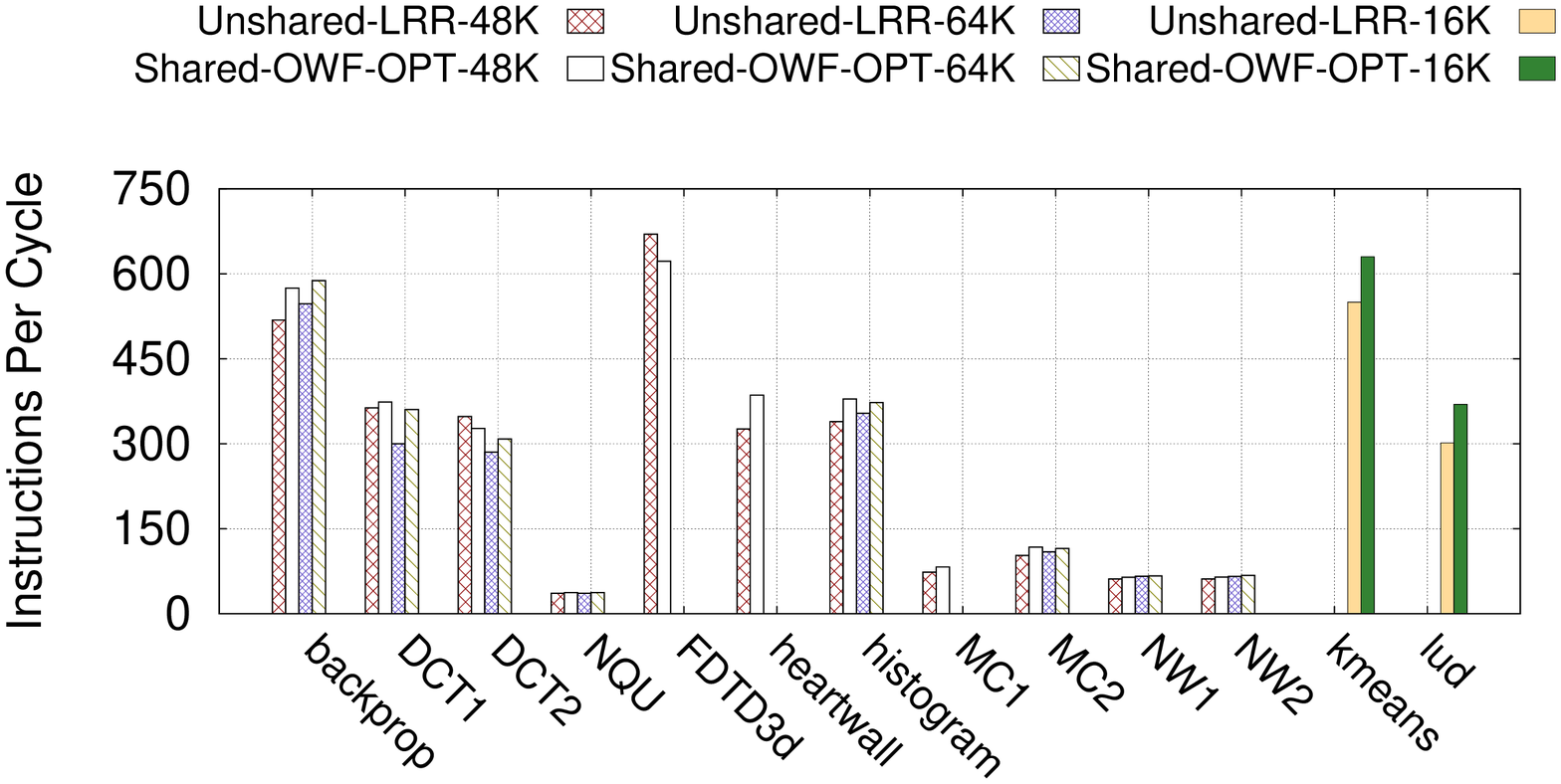}
\vskip -2mm
\caption{Performance Analysis  for Various Configurations}\label{fig:ipc_new}
\vskip -3mm
\end{figure}

\begin{figure}
\centering
\includegraphics[scale=0.4]{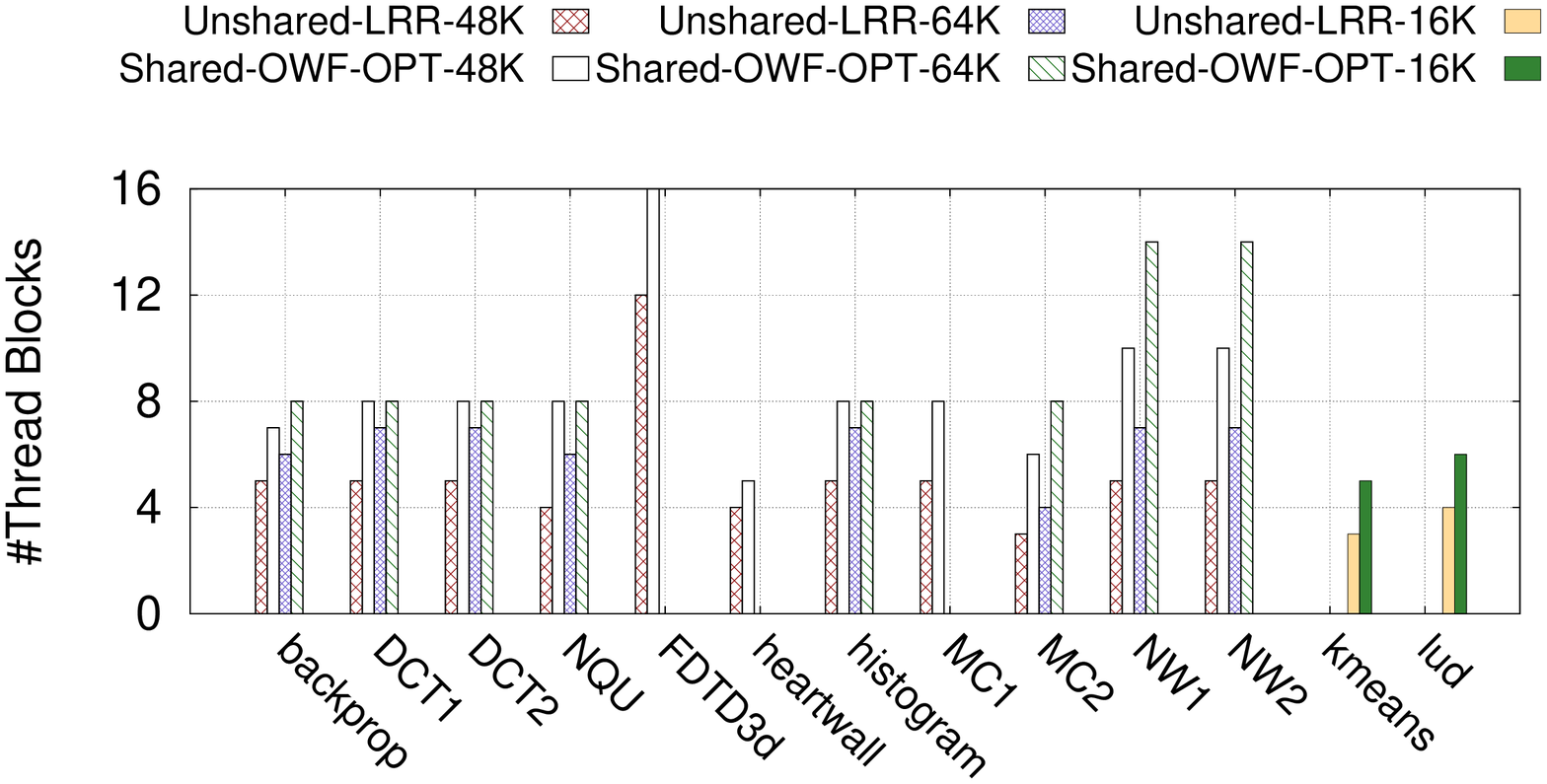}
\vskip -2mm
 \caption{No. of Thread Blocks for Various Configurations}\label{fig:blocks_new}
\vskip -3mm
\end{figure}

\textit{Increase in Number of Resident Thread Blocks:}
Figure~\ref{fig:blocks_new} shows the increase in number of
resident thread blocks for scratchpad sharing. In the figure, we
use \emph{Unshared-LRR-48K} to denote the baseline approach that
uses 48KB scratchpad memory (according to Configuration-1) and
\emph{Shared-OWF-OPT-48K} to denote the scratchpad sharing
approach (with all optimizations) that use 48KB scratchpad
memory. The notations \emph{Unshared-LRR-64K} and
\emph{Shared-OWF-OPT-64K}, which use 64KB scratchpad memory
configuration, are defined analogously. From the figure, we
observe that all the applications launch more number of resident
thread blocks with our approaches (\emph{Shared-OWF-OPT-48K} and
\emph{Shared-OWF-OPT-64K}) when compared to their respective
baseline configurations. The applications \emph{DCT1}, \emph{DCT2}, and
\emph{FDTD3d} launch 8, 8, and 16 thread blocks respectively with
\emph{Shared-OWF-OPT-48K}, which is the limit on the number of
resident threads in the SM. Similarly \emph{backprop},
\emph{DCT1}, and \emph{DCT2} launch 8 thread blocks each with
\emph{Shared-OWF-OPT-64K}, which is also the limit on the number
of resident threads in the SM. For all the other applications,
our approach is able to launch maximum number of thread blocks
that is possible with scratchpad sharing approach. Hence,
scratchpad sharing helps in increasing the number of thread
blocks for the applications even with the configurations that use
larger scratchpad memory per SM.

Figure~\ref{fig:ipc_new} shows the performance comparison of our
approach with the two baseline configurations.  We observe that with 
\emph{Shared-OWF-OPT-48K}, all the applications
except \emph{FDTD3d} and \emph{DCT2} show performance improvement
when compared to \emph{Unshared-LRR-48}.  Similarly with
\emph{Shared-OWF-OPT-64K}, all the applications perform better
when compared to \emph{Unshared-LRR-64K}. Consider the
applications \emph{backprop}, \emph{DCT1}, \emph{histogram},
\emph{MC2}, and \emph{NQU}. These applications with
\emph{Shared-OWF-OPT-48K} perform better even when compared to
\emph{Unshared-LRR-64K}.  Also, these applications, when used
with \emph{Shared-OWF-OPT-64K}, perform better than
\emph{Unshared-LRR-64K} as well. \emph{NW1} and \emph{NW2} show
improvement with scratchpad sharing when compared to their
respective baseline approaches. Applications \emph{heartwall} and \emph{MC1},
where scratchpad sharing is applicable only with Configuration-1
(Table~\ref{table:KeplerMaxwell}), show improvement when compared
to \emph{Unshared-LRR-48K}. For \emph{DCT2}, increasing the
amount of scratchpad memory from 48KB to 64KB per SM does not
improve the performance of \emph{Unshared-LRR} configuration
since it increases number of stall cycles in the SM. Hence
increase in the number of thread blocks for
\emph{Shared-OWF-OPT-48K} does not improve its performance w.r.t
\emph{Unshared-LRR-48K}. However, \emph{Shared-OWF-OPT-48K}
performs better than \emph{Shared-OWF-OPT-64K} because \emph{Shared-OWF-OPT-48K} has
lesser number of thread blocks that make more progress, hence
reducing resource contention. \emph{FDTD3d} does not show
improvement with scratchpad sharing due to increase in the stall
cycles with our approach. Also additional benchmarks,
\emph{kmeans} and \emph{lud} show improvement with
\emph{Shared-OWF-OPT-16K} when compared to
\emph{Unshared-LRR-16K}.

To summarize, our approach helps
in improving the performance of the applications even  with increase  in the  size  of the  scratchpad memory  per
SM.

\begin{table}[t]
  \caption{Benchmarks used for comparison with Shared Memory Multiplexing~\cite{SharedMemMultiplexing}\label{table:set_5}}
  \vskip -2mm 
  \scalebox{0.9}{\renewcommand{\arraystretch}{1.0}
    \begin{tabular}{l@{\ }l@{\ }l@{\ }l@{\ }l@{\ } l@{}}
      \hline\hline
      Application & Kernel  & \#Scratchpad & Scratchpad   & Block \\
      &         & Variables    &  Size (Bytes)& Size  \\
      \hline
      Convolution (CV) & convolutionColumnsKernel  & 1 & 8256 & 128\\
      Fast Fourier Transform (FFT)   & kfft & 1 & 8704 &  64 \\
      Histogram (HG) & histogram256 & 1 & 7168   &32 \\
      MarchingCubes (MC) & generateTriangles & 2 & 9216 & 32\\
      Matrix Vector Multiplication (MV) & mv\_shared  & 1 & 4224 & 32\\
      ScalarProd (SP) & scalarProdGPU & 1 & 4114 & 64\\
      \hline
      \end{tabular}}
\end{table}
\begin{table}[t]
\centering
\caption{Details of Modifications to the Shared Memory Multiplexing Benchmarks} \vskip -3mm
\label{tab:modifications_mux}
{\scalebox{0.88}{\renewcommand{\arraystretch}{1.05}
 \begin{tabular}{l@{\ \ }l@{\ \ }l@{\ }l@{\ }l@{\ \ } l@{}}
\hline \hline
Application & File Name & Line  & \begin{tabular}[c]{@{}l@{}}Variable Name\\ \end{tabular} & Original  & New  \\ 
& & Number & & Configuration & Configuration \\
\hline
MC & marchingCubes.cpp &  202 & g\_bQAReadback & flase & true \\ \hline
CV & main.cpp &  73 & iterations & 10 & 1 \\ \hline
SP & scalarProd\_kernel.cu &  42 & accumResult (size) & ACCUM\_N & ACCUM\_N+12 \\ \hline
HG & hist.cu & \begin{tabular}[c]{@{}l@{}}93 \\ 43 \end{tabular} & \begin{tabular}[c]{@{}l@{}}iterator \\ - \end{tabular} & \begin{tabular}[c]{@{}l@{}}16\\ runTest(1024, 1024) \end{tabular} & \begin{tabular}[c]{@{}l@{}}1 \\ runTest(1024, 896);\end{tabular} \\ 
   & hist\_kernel.cu & \begin{tabular}[c]{@{}l@{}}3\\ 4\end{tabular} & \begin{tabular}[c]{@{}l@{}}BIN\_SIZE \\ THREAD \end{tabular} & \begin{tabular}[c]{@{}l@{}}128\\  64 \end{tabular} &  \begin{tabular}[c]{@{}l@{}}224\\ 32\end{tabular} \\ \hline
MV & mv.cu & 107 & iterator & 16  & 1  \\ \hline
\end{tabular}}}
\vskip -2mm
\end{table}

\begin{figure}
\centering
\includegraphics[scale=0.4]{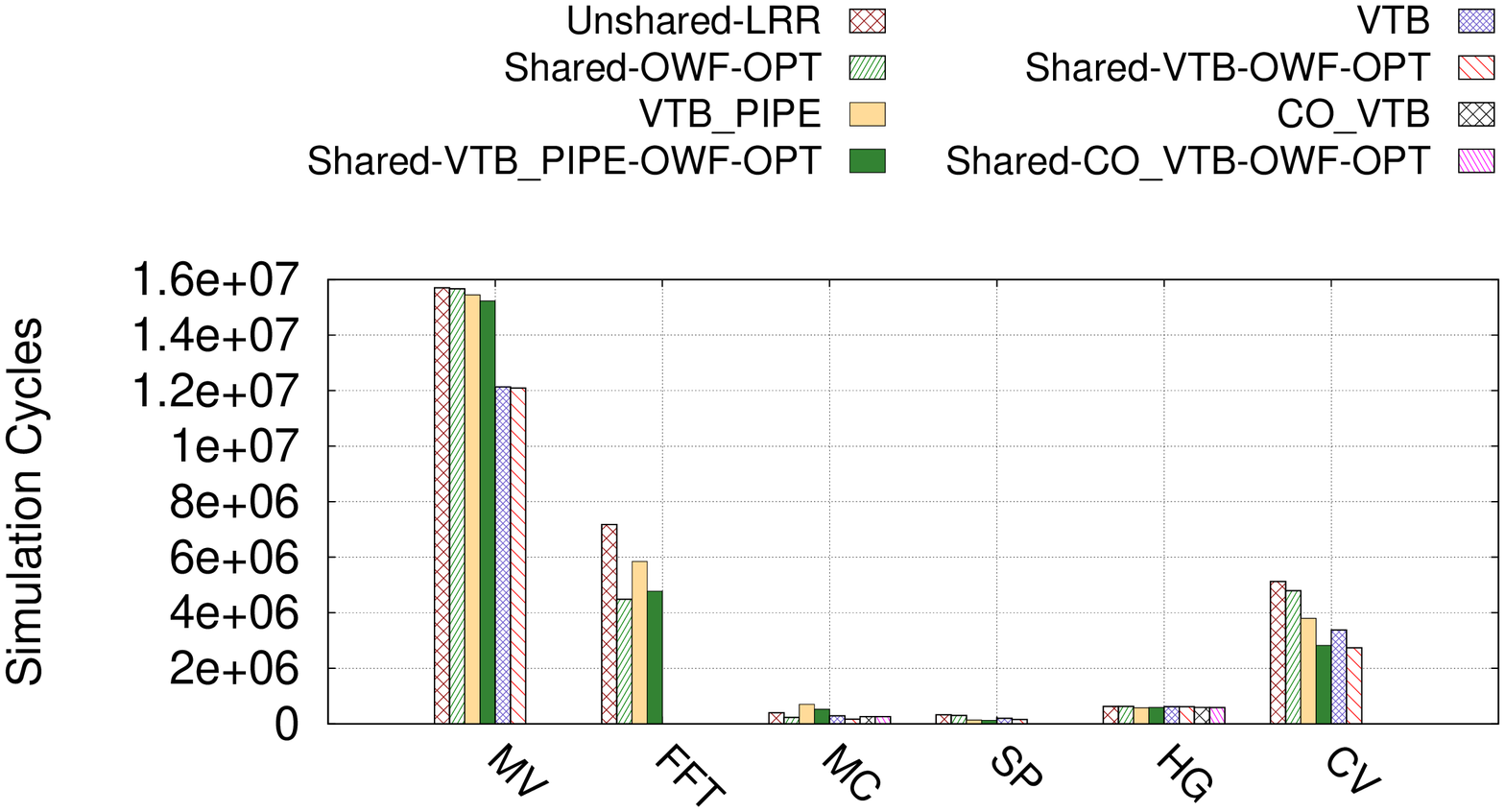}
\vskip -2mm
\caption{Performance Comparison with Other Approaches in terms of Number of Simulation Cycles}\label{fig:mux_compare_simcycles}
\vskip -3mm
\end{figure}  

\begin{table}[t]
  \caption{Comparing the Number of Simulated Instructions for Various Approaches}  
\vskip -2mm 
\centering
\scalebox{0.77}{\renewcommand{\arraystretch}{1.05}
\begin{tabular}{@{}l@{ }r@{ }r@{\ }r@{\ }r@{\quad}r@{}@{\ }r@{}@{\ }r@{}@{\ }r@{}}
\hline\hline
Benchmark & Unshared- & Shared- & VTB\_PIPE & Shared-VTB\_ & VTB & Shared-VTB\_ & CO\_VTB & Shared-CO\_ \\
& LRR&  OWF-OPT & & PIPE-OWF-OPT & & OWF-OPT & & VTB-OWF-OPT \\ 
\hline
MV & 412,680,192 & 412,712,960 & 637,648,896 & 637,698,048 & 696,909,824 & 696,942,592 & N/A & N/A \\
FFT & 416,022,528 & 416,284,672 & 445,644,800 & 446,038,016 & N/A & N/A & N/A & N/A \\
MC & 24,321,552 & 24,387,088 & 29,228,560 & 29,294,096 & 27,352,592 & 27,418,128 & 25,456,176 & 25,521,712 \\
SP & 27,472,640 & 27,505,408 & 15,611,648 & 15,644,416 & 15,946,752 & 15,979,520 & N/A & N/A \\
HG & 19,898,368 & 19,906,560 & 10,682,368 & 10,690,560 & 10,430,464 & 10,438,656 & 13,253,518 & 13,261,710 \\
CV & 439,812,096 & 440,991,744 & 525,729,792 & 527,499,264 & 497,025,024 & 498,794,496 & N/A & N/A \\
\hline
\end{tabular}}
\label{table:insn_compare}
\vskip -2mm
\end{table}

\subsubsection{Performance Comparison with Shared Memory Multiplexing~\cite{SharedMemMultiplexing}} ~\label{sec:compare}  

\begin{figure}
\centering
\includegraphics[scale=0.4]{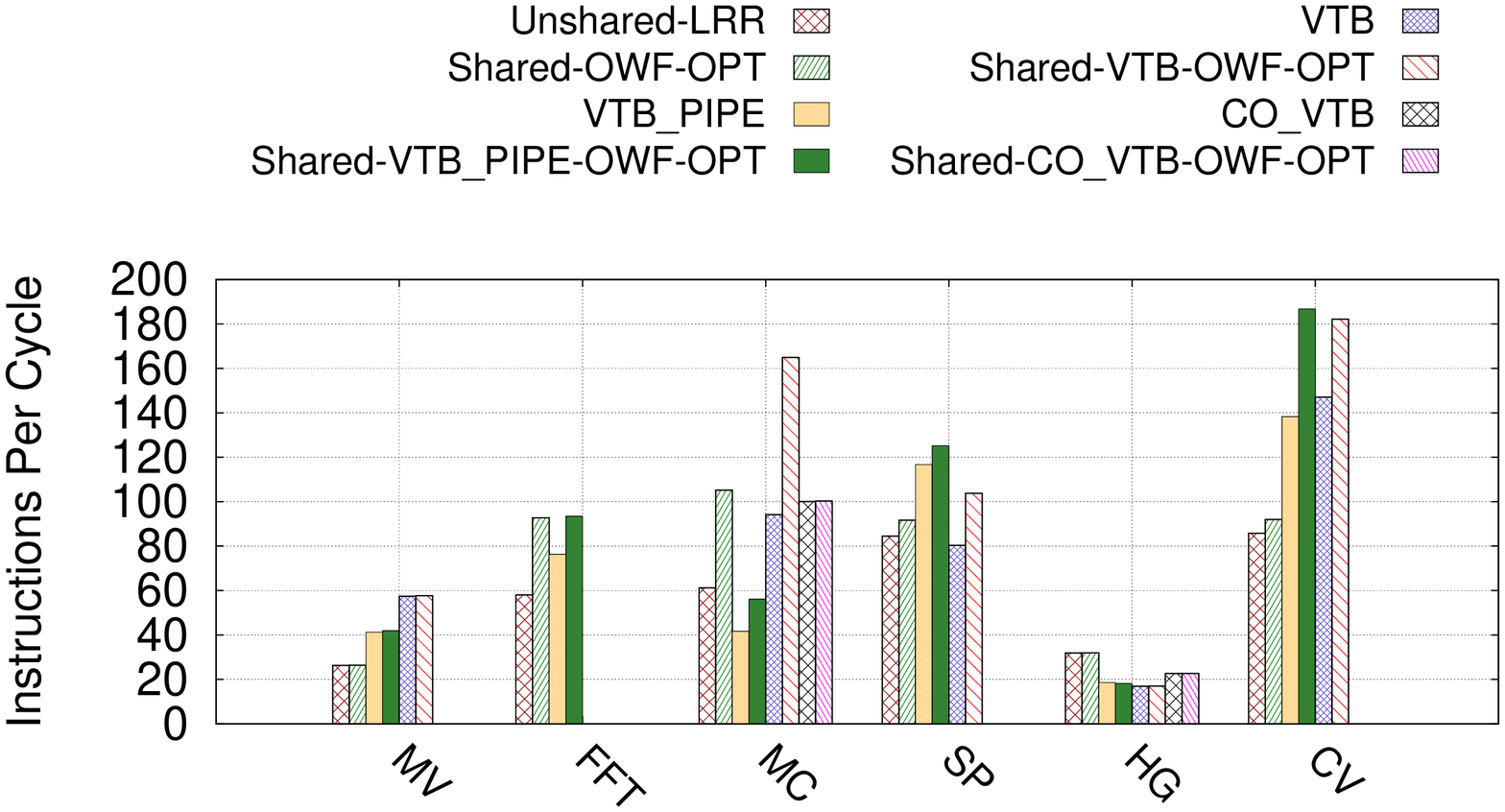}
\vskip -2mm
\caption{Performance Comparison with Other Approaches in terms of
IPC }\label{fig:mux_compare_ipc}
\vskip -3mm
\end{figure}

Figures~\ref{fig:mux_compare_simcycles} and \ref{fig:mux_compare_ipc} compare the performance of
  our approach with the software approaches proposed
  by~\citeN{SharedMemMultiplexing} in terms of number of simulation cycles and instructions
  per cycles respectively.
  We use their benchmarks
  (Table~\ref{table:set_5}), and simulate them on the GPU
  configuration shown in Table~\ref{table:GPGPUArch}. In the
  figure, \emph{Unshared-LRR} and \emph{Shared-OWF-OPT} denote
  the baseline and scratchpad sharing approaches
  respectively. \emph{VTB}, \emph{VTB\_PIPE}, and \emph{CO\_VTB}
  denote the compiler optimizations proposed
  by~\citeN{SharedMemMultiplexing}\footnote{ Note that, as
    described in~\citeN{SharedMemMultiplexing}, \emph{CO\_VTB} is
    suitable only for few workloads (i.e., \emph{MC} and
    \emph{HG}). Also, for \emph{FFT}, we do not compare
    \emph{VTB} with \emph{VTB\_PIPE} because \emph{VTB} combines
    4 thread blocks whereas \emph{VTB\_PIPE} combines 2 thread
    blocks in their implementation.}.  Similarly, we use
  \emph{Shared-VTB-OWF-OPT}, \emph{Shared-VTB\_PIPE-OWF-OPT}, and
  \emph{Shared-CO\_VTB-OWF-OPT} to measure performance of
  scratchpad sharing on the applications that are optimized with
  \emph{VTB}, \emph{VTB\_PIPE}, and \emph{CO\_VTB} respectively.

From Figure~\ref{fig:mux_compare_simcycles} we observe that the application \emph{MC} performs better (spends less number of
simulation cycles) with
\emph{Shared-OWF-OPT} than with \emph{Unshared-LLR}, \emph{VTB},
\emph{VTB\_PIPE}, and \emph{CO\_VTB} approaches.  Interestingly,
applying \emph{Shared-OWF-OPT}  on top of \emph{VTB},
\emph{VTB\_PIPE}, or \emph{CO\_VTB} improves the performance further. Similarly,
\emph{FFT} shows improvement with \emph{Shared-OWF-OPT} when
compared to \emph{Unshared-LRR} and \emph{VTB\_PIPE}. In this case also
\emph{Shared-VTB\_PIPE-OWF-OPT} outperforms {VTB\_PIPE}. 
In contrast, for  the application \emph{HG}, sharing does not impact 
the performance, even on the top of \emph{VTB} and \emph{VTB\_PIPE} 
optimizations. This is because  the additional thread blocks launched do not 
make much progress before they start accessing shared scratchpad.
For the same reason, scratchpad sharing does not have any further impact 
on \emph{MV}.  The applications \emph{CV} and \emph{SP} perform better 
with \emph{VTB\_PIPE} than with \emph{Shared-OWF-OPT}. However, 
the performance is further improved when scratchpad sharing is 
combined with \emph{VTB} and \emph{VTB\_PIPE} approaches.
We also observe a change in the number of executed instructions 
for \emph{VTB}, \emph{VTB\_PIPE}, and \emph{CO\_VTB} approaches
for all the applications (shown in Table~\ref{table:insn_compare}) when compared the \emph{Unshared-LRR}, because they modify the benchmarks. 

It can be concluded from these experiments that scratchpad
sharing and shared memory multiplexing approaches compliment each
other well, and most applications show the best performance when
the two approaches are combined.

\subsubsection{Performance Analysis for Different Number of SMs} ~\label{sec:varysms}  
\begin{figure}
\centering
\includegraphics[scale=0.53]{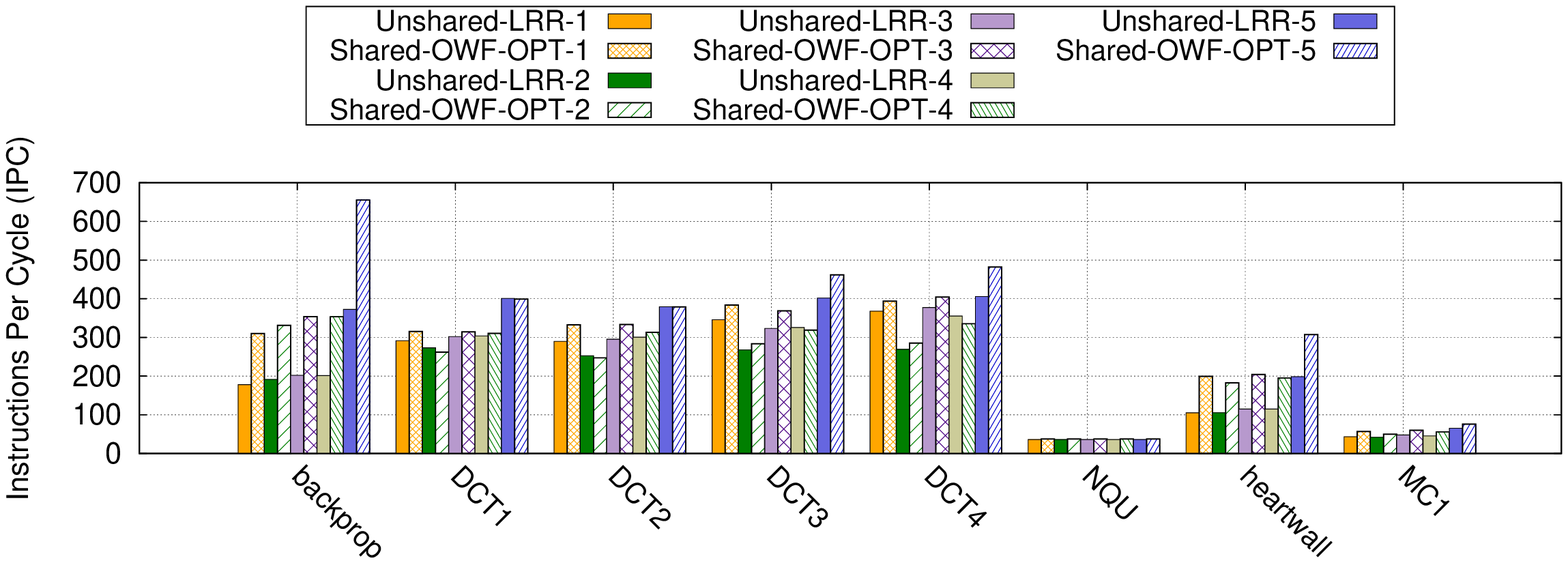}
\vskip 1mm
\caption{Performance Analysis for Different Number of SMs}
\label{fig:varysm}
\vskip -7mm
\end{figure}

Figure~\ref{fig:varysm} shows the effectiveness of our approach when different number of SMs are used in GPUs. Table~\ref{table:varysms} shows five different configurations of SMs (obtained from \citeN{GPGPUSIM}) that specify the number of clusters and number of SMs in each cluster. In the figure \emph{Unshared-LRR-1} and \emph{Shared-OWF-OPT-1} represent the baseline and scratchpad sharing approaches when they are simulated with \emph{ SM Config-1}. Similarly, we use other notations to represent the approaches for other configurations.

From the figure, we observe that applications show maximum benefit with baseline approach when the number of SMs is 30. However, the applications show further improvements when scratchpad sharing is applied on them. Application \emph{heartwall} shows maximum benefit with scratchpad sharing, because the additional thread blocks launched in our approach do not access shared scratchpad memory. For \emph{backprop}, the performance improves with increase in the number of SMs, since increase in the number of SM leads to increase in TLP. The performance improves further when scratchpad sharing is applied. \emph{MC1} also shows improvement with scratchpad sharing for all configurations when compared to baseline, because the additional thread blocks launched in our approach make significant progress before they start accessing shared scratchpad memory. \emph{DCT1, DCT2, DCT3, and DCT4} perform better when the number of SMs is 14 (\emph{SM Config-1}) when compared to 15 (\emph{SM Config-2}). This is because \emph{SM Config-2} contains more number of SMs per clusters than \emph{SM Config-1}, and all the SMs within a cluster share a common  port with interconnect. \emph{NQU} shows similar improvements for all configurations with our approach. Because even with increase in the number of SMs, baseline approach does not show improvement. 

From the results we conclude that, scratchpad sharing helps in improving the performance applications even by varying number of SMs in a GPU. The applications improve further when there are more number of SMs with less number of SMs within a cluster in GPUs.

\begin{table}[t]
  \caption{GPGPU-Sim Configurations with Various Number of SMs}\label{table:varysms}
  \vskip -2mm 
  \scalebox{0.86}{\renewcommand{\arraystretch}{1.0}\begin{tabular}{l|r|r|r|r|r}
  
    \hline\hline
    Resource/Core &  SM Config-1 &   SM Config-2 & SM Config-3  & SM Config-4 &  SM Config-5 \\          
    \hline
    Total Number of SMs &  14	& 15 & 16 & 16 & 30 \\
    Number of Cluster * Number  & 7*2 & 3*5 & 8*2 & 4*4 & 10*3 \\
    of SMs per Cluster &  & & & & \\
    \hline
  \end{tabular}}
\end{table}

\section{Related Work} \label{sec:relatedwork}
Resource sharing technique~\cite{resourcesharing}, proposed by the
authors earlier, improves the throughput by minimizing the register
and scratchpad memory underutilization by modifying the GPU
architecture and scheduling algorithm.  This work improves on
it by introducing compiler optimizations for better layout of
scratchpad variables and early release of shared
scratchpad. Other related approaches that improve
the performance of GPUs are discussed below:

\subsection{Resource Management in GPUs}
  Shared memory
  multiplexing~\cite{SharedMemMultiplexing} proposes solutions
  that come closest to our approach.  They provide software and
  hardware solutions to address the TLP problem caused by limited
  shared memory.  The software approach combines two thread
  blocks into a single virtual thread block. The two thread
  blocks in a virtual block can execute instructions in parallel,
  as long as they do not access shared memory; and become serial
  when they need to access shared memory.  The paper also
  describes a mechanism (called CO-VTB) that divides the shared
  memory into private and public part so that the thread blocks
  in a virtual block can access the private part in parallel and
  the public part in serial. However, CO-VTB has a high overhead
  of partitioning the data into private and public part, and is
  not suitable for all workloads. Also, they need to generate the
  code manually.  The paper also gives a hardware solution to
  dynamically allocate and deallocate scratchpad memory using the
  existing barrier instruction.  Again, these instructions need
  to be inserted manually in the code, and nesting of the barrier
  instructions is not allowed in order to avoid any deadlocks.

  In contrast, ours is a hardware solution that allows launching
  additional thread blocks in each SM. These additional thread
  blocks use the wasted scratchpad memory, and also share part of
  the allocated scratchpad memory with other resident thread
  blocks. The additional thread blocks launched in our approach
  make progress as long as they do not require shared scratchpad
  memory, and wait until the shared scratchpad is released by
  other thread blocks. Our approach is fully automatic---our
  compiler optimization automatically identifies the regions of
  the shared and unshared scratchpad memory at run-time and
  inserts instruction to release the shared scratchpad as early
  as possible.  Even in the presence of barrier instructions, our
  approach can not have deadlocks. In addition, we propose a warp
  scheduling mechanism that effectively schedule these additional
  warps to hide the long latencies in a better way. 

  Warp level divergence technique~\cite{WarpLevelDivergence}
  improves the TLP by minimizing register underutilization.  It
  launches one additional partial thread block when there are
  insufficient number of registers for an entire thread block.
  However, the number of warps in the partial thread block is
  decided by the number of unutilized registers, and also the
  partial thread block does not share registers with any other
  thread blocks.  The unified storage approach~\cite{Unification}
  allocates the resources of SM (such as registers, scratchpad
  memory, and cache) dynamically as per the application demand.
  \citeN{ondemand} use virtual registers to launch more thread
  blocks. These registers are mapped to the physical registers as
  per the demand. Our compiler optimizations can help in early
  release of unused registers with this approach.  
  \citeN{lockscratchpad} 
  describe a mechanism to lock and unlock parts of scratchpad memory.
  We can reutilize the existing mechanism by defining a custom 
  hash function that maps shared scratchpad memory regions to
  corresponding lock addresses. For unshared scratch region, 
  the access can be given directly (i.e., the lock can be 
  granted always).
  \citeN{NMNL}
  propose a dynamic algorithm to launch the optimal number of
  thread blocks in an SM to reduce the resource contention. We
  can combine their techniques with our approach to reduce the
  number of increase in the stall cycles can occur with the
  additional thread blocks.  \citeN{virtual} propose a resource
  virtualization scheme for sharing of GPU resources with
  multiprocessors. The virtualization layer proposed by them
  helps in improving the performance by overlapping multiple
  kernels executions.  

  Our approach is different from the above in that it addresses
  underutilization of scratchpad memory by launching more than
  one additional thread blocks in each SM. These thread blocks
  share scratchpad memory with other resident thread blocks, thus
  improve its utilization. The proposed OWF-scheduler reduces
  stall cycles and improves the performance further.
\vspace*{-2mm}

\subsection{Compiler Optimizations for Efficient Resource Utilization in GPUs}

\citeN{Integer} formulated the problem of scratchpad memory allocation
as an integer programming problem, which maximizes scratchpad memory
access and minimizes device memory access to improve GPU
performance. Their framework can allocate parts of arrays on
scratchpad, and also suggest profitable loop transformations.
\citeN{UnifiedOnChip} proposed on-chip memory allocation scheme for
efficient utilization of GPU resources.  It aims to alleviate register
pressure by spilling registers to scratchpad memory instead of local
memory. \citeN{CRAT} proposed a compile time coordinated register
allocation scheme to minimize the cost of spilling registers. These
schemes do not propose any architectural change to GPUs and are
orthogonal to our approach of scratchpad sharing.

\subsection{Scheduling Techniques for GPUs} 

The two level warp scheduling algorithm, proposed by \citeN{TwoLevel}, 
forms groups of warps and uses LRR to schedule warps in a group.
It also proposes a large warp microarchitecture to minimize
resource underutilization.  \citeN{CAWS} hide the long execution
latencies by scheduling critical warps more frequently than other than
warps. It helps in finishing the thread block sooner thus improving
resource utilization.  However, it requires the knowledge of critical
warps. To address the problem, \citeN{CAWA} proposed a coordinated
solution that identifies the critical warps at run time using
instructions and stall cycles. Further, they proposed a greedy based
critical warp scheduling algorithm to accelerate the critical warps in
the SMs. OWL~\cite{OWL} provides a scheduling mechanism to reduce
cache contention and to improve DRAM bank level
parallelism. \citeN{ThreadBlockScheduling} focus on reducing resource
contention by providing lazy thread block scheduling mechanism. They
also proposed block level CTA scheduling policy that allocates
consecutive CTAs into the same SM to exploit cache locality. Their
approach can also be integrated to our approach.

\subsection{Improving GPU Performance through Memory Management}
Several other approaches exploit memory hierarchy to improve the
performance of GPU applications.  \citeN{AutoOnChip} proposed compiler
techniques to efficiently place data onto registers, scratchpad
memory, and global memory by analyzing data access
patterns. \citeN{mascar} proposed a scheduling policy that improves
the GPU performance by prioritizing memory requests of single warp
when memory saturation occurs. \citeN{PriorityCache} provide a
mechanism to handle the cache contention problem that occurs due to
increased number of resident threads in an SM. Their approach is
alternative to the earlier proposed thread throttling
techniques~\cite{DivergenceAware,Rogers,NMNL}.

\subsection{Problems with Warp Divergence}
Other techniques to improve GPU performance is by handling warp
divergence.  Dynamic warp formation~\cite{DynWarpFormation} addresses
the limited thread level parallelism that is present due to branch
divergence. It dynamically forms new warps based on branch target
condition. However, the performance of this approach is limited by the
warp scheduling policy.  Thread block compaction~\cite{TBCompaction}
addresses the limitation of dynamic warp formation that occurs when
the new warps that are formed may require more number of memory
accesses. Their approach provides a solution by regrouping the new
warps at the reconverging points. However in their solution, warps
need to wait for other warps to reach the divergent path. \citeN{Linearization} proposed linearization technique to avoid
duplicate execution of instructions that occurs due to branch
divergence in GPUs. \citeN{Brunie,Han} provide hardware and
software solutions to handle branch divergence in GPUs.


\subsection{Miscellaneous}

Warped pre-execution~\cite{preExecution} accelerates a single warp by
executing independent instructions when a warp is stalled due to long
latency instruction.  It improves the GPU performance by hiding the
long latency cycles in a better way.  \citeN{AffineLoops} proposed a
compiler framework for optimizing memory access in affine loops.
\citeN{reductions,FFT} show that several applications are
improved by using scratchpad memory instead of using global memory.

\section{Conclusions and Future Work} \label{sec:conclusion}

In this paper, we propose architectural changes and compiler
optimizations for sharing scratchpad effectively to address the
underutilization of scratchpad memory in GPUs.  Experiments with
various benchmarks help us conclude that if the number of resident
thread blocks launched by an application are limited by scratchpad
availability (Table~\ref{table:set_1}), scratchpad sharing (with the
compiler optimizations) improves the performance.  On the other hand,
for other applications where the number of thread blocks is not
limited by scratchpad availability (Table~\ref{table:set_3}), the
hardware changes do not negatively impact the run-time.

In future, we would like to extend our work to integrate register
sharing approach~\cite{resourcesharing}. Value range analysis
techniques~\cite{RangeAnalysis,BufferOverflow}, typically employed for
detecting buffer overflows, can be incorporated in our approach to
refine the access ranges of shared scratchpad variables, thus help
release shared scratchpad even earlier. We need to study the
impact of hardware changes on power consumption, and find ways to minimize it.

\begin{acks}
We thank the anonymous reviewers for their useful feedback. We
  also thank the authors of ``Shared Memory Multiplexing: A Novel
  Way to Improve GPGPU Throughput''~\cite{SharedMemMultiplexing}
  for sharing their benchmarks with us.  
\end{acks}

\bibliographystyle{ACM-Reference-Format-Journals}
\bibliography{CompOptJournal}

\clearpage
\appendix
\section{Appendix}\label{sec:appendix}

\begin{table}[h]
  \caption{Comparing IPC values of Scratchpad sharing with Various Baseline Implementations}  
\vskip -2mm 
\centering
\scalebox{1.0}{\renewcommand{\arraystretch}{1.05}
\begin{tabular}{@{}l@{ }r@{ \quad}r@{\quad}r@{\quad}r@{}}
\hline\hline
Benchmark & Unshared-LRR & Unshared-GTO & Unshared-2Level & Shared-OWF-OPT \\
\hline
backprop & 178.01 & 179.78 & 178.38 & 310.1 \\
DCT1 & 284.48 & 289.39 & 281.54 & 322.28 \\
DCT2 & 283.84 & 287.16 & 288.42 & 325.83 \\
DCT3 & 358.11 & 373.21 & 384.3 & 423.12 \\
DCT4 & 381.23 & 402.4 & 406.12 & 436.2 \\
NQU & 35.77 & 35.77 & 35.76 & 37.46 \\
SRAD1 & 199.18 & 200.58 & 199.23 & 227.74 \\
SRAD2 & 67.19 & 67.39 & 66.98 & 76.18 \\
FDTD3d & 330.52 & 331.96 & 328.95 & 322.94 \\
heartwall & 104.92 & 104.92 & 105.05 & 201.62  \\
histogram & 153.46 & 153.38 & 151.56 & 153.19 \\
MC1 & 44.43 & 44.57 & 44.15 & 58.79 \\
NW1 & 25.34 & 25.34 & 25.26 & 25.94 \\
NW2 & 25.4 & 25.4 & 25.32 & 27.51 \\
\hline
\end{tabular}}
\label{table:ipc}
\vskip -2mm
\end{table}

In Figure~\ref{fig:ipc}, we have shown the performance of scratchpad sharing approach when normalized with respect to \emph{Unshared-LRR}. In Table~\ref{table:ipc}, we show the absolute number of instructions executed per cycle (IPC) for scratchpad sharing approach, and we compare it with that of baseline implementation that uses LRR, GTO, and two-level scheduling policies.


\medskip

\end{document}